\documentclass[aps,prb,twocolumn,amsmath,amssymb]{revtex4}

\usepackage{graphicx}
\usepackage{dcolumn}
\usepackage{bm}
\usepackage[caption=false]{subfig}
\usepackage[symbol]{footmisc}
\usepackage{amssymb}
\usepackage{booktabs}
\usepackage{amsmath}
\usepackage{graphicx}
\usepackage{verbatim}
\usepackage{tabularx}
\usepackage[usenames, dvipsnames]{color}
\usepackage{hyperref}
\usepackage{color,soul}

\def\beq{\begin{equation}}
	\def\eeq{\end{equation}}
\def\vk{{\bf k}}
\renewcommand{\thefootnote}{\fnsymbol{footnote}}

\begin{document}
	
\title{High-order crystal field and rare-earth magnetism in RECo$_5$ intermetallics }
\author{L. V. Pourovskii$^{1,2}$, J. Boust$^1$, R. Ballou$^3$, G. Gomez Eslava$^3$, D. Givord$^3$\footnote[2]{Deceased 4 February 2019}}

	\affiliation{
		$^1$Centre de Physique Th\'eorique, Ecole Polytechnique, CNRS, Institut Polytechnique de Paris, 91128 Palaiseau Cedex, France \\
		$^2$Coll\`ege de France, 11 place Marcelin Berthelot, 75005 Paris, France \\
		$^3$CNRS, Universit\'e Grenoble Alpes, Grenoble INP, Institut N\'eel, 38000 Grenoble, France}

\begin{abstract}
	Crystal-field (CF) effects on the rare-earth (RE) ions in ferrimagnetic intermetallics NdCo$_5$ and TbCo$_5$ are evaluated using an {\it ab initio} density functional + dynamical mean-field theory approach in conjunction with a quasi-atomic approximation for on-site electronic correlations on the localized 4$f$ shell. The study reveals an important role of the high-order sectoral harmonic component of the CF in the magnetism of RECo$_5$ intermetallics. 
	An unexpectedly large value is computed in the both systems for the corresponding crystal-field parameter (CFP) $A_6^6 \langle r^6 \rangle$, far beyond what one would expect from only electrostatic contributions. It allows solving the enigma of the non-saturation of zero-temperature Nd magnetic moments in NdCo$_5$ along its easy axis in the Co exchange field. This unsaturated state had been previously found out from magnetization distribution probed by polarised neutron elastic scattering but had so far remained theoretically unexplained. The easy plane magnetic anisotropy of Nd in NdCo$_5$ is strongly enhanced by the large value of $A_6^6\langle r^6 \rangle$. Counter-intuitively, the polar dependence of anisotropy energy within the easy plane remains rather small. The easy plane magnetic anisotropy of Nd is reinforced up to high temperatures, which is explained through $J$-mixing effects. The calculated {\it ab initio} anisotropy constants of NdCo$_5$ and their temperature dependence are in quantitative agreement with experiment. Unlike NdCo$_5$, the $A_6^6 \langle r^6 \rangle$ CFP has negligible effects on the Tb magnetism in TbCo$_5$ suggesting that its impact on the RE magnetism is ion-specific across the RECo$_5$ series. The origin of its large value is the hybridization of RE and Co states in a hexagonally coordinated local environment of the RE ion in RECo$_5$ intermetallics.
\end{abstract}

\maketitle
\renewcommand*{\thefootnote}{\arabic{footnote}}

\section{Introduction}

Magnetic properties of transition-metal (TM)-rare-earth (RE) intermetallics are determined by a subtle interplay between metallic TM $d$ electrons and ionic RE $f$ electrons. Among those apt at giving rise to permanent magnets \cite{Strnat1991,Buschow1991,Hirosawa2017} the TM constituent is a late 3$d$ TM, such as Fe or Co, providing a large magnetization and a high Curie temperature that can reach 1000~K. The RE magnetism in these intermetallics is essentially induced by an exchange field due to the TM ferromagnetic order. The direct exchange coupling between the RE 4$f$ magnetic moments is comparatively much weaker \cite{Nicklow1976,Fuerst1986,Kuzmin2004} and can be neglected. The magnetic anisotropy qualifying such magnets as hard arises, on the other hand, dominantly from the RE sublattice, especially at low and intermediate temperatures. It stems from the strong spin-orbit (SO) coupling at the 4$f$ shell transferring  to magnetism the anisotropy of crystal-field (CF). The magnitude of this RE single-ion anisotropy (SIA) is thus determined by the CF acting on the 4$f$ shell and its interplay with the TM-induced exchange field $B_{\mathrm{ex}}$ of a comparable magnitude.


The so-called two-sublattice model shortly outlined above is believed to be relevant to the  RECo$_5$\cite{Strnat1991}, RE$_2$Co$_{17}$, and RE$_2$Fe$_{14}$B \cite{Buschow1991,Herbst1991} material families, which comprise key modern high-performance magnets. Among the quantities determining the RE SIA in these materials, i.~e. the TM magnetization, TM-RE exchange coupling and CF\cite{Kuzmin2007}, the later is particularly hard to assess both experimentally and theoretically. In particular, extracting CF parameters (CFPs) from high-field magnetization measurements (see e.~g. Refs~\onlinecite{Klein1975,Tie1991,Bartashevich1993,Bartashevich1994,Zhang1994,Kostyuchenko2015}) is subject to significant uncertainties, as the total magnetization and macroscopical anisotropy constants measured in such experiments should be subsequently separated into the RE and TM contributions on the basis of a particular microscopical two-sublattice model. In the analysis of such experiments it is usual to include only low-rank CFPs and to restrict the consideration to the ground-state (GS) RE multiplet \cite{Buschow1974,Radwanski1986,Tie1991}. The parameter-free {\it ab initio} prediction of RE CFPs is a notoriously difficult problem, mainly due to inability of the conventional density functional theory (DFT)  to correctly account for the physics of localized 4$f$  shells. The standard DFT-based approach, extensively applied to RE-TM intermetallics\cite{Novak1994,Hummler1996,Novak1996,Divis2005,Novak_book2013,Novak2014c,Harashima2016}, is to treat RE 4$f$s as an "open-core" shell , meaning that their hybridization with other valence states is completely neglected. The validity of this "open-core" approximation for the CF in real TM-RE intermetallics  is usually hard to assess because of the above-mentioned uncertainties in extracting RE CFPs from magnetization measurements.

The RECo$_5$ family represents a suitable testbed for theoretical approaches to CF effects in TM-RE intermetallics. This family crystallizes in a simple hexagonal structure within which the RE ions are distributed over a single site. The magnetic behavior of RECo$_5$ exhibits a rich variety along the series: SmCo$_5$ features a very strong uniaxial anisotropy being the first widely used  RE-based permanent magnet\cite{Strnat1991}. On the other hand, with RE = Nd, Tb, and Dy the low-temperature magnetic anisotropy of RECo$_5$ is of an easy-plane type. When the temperature is increased these intermetallics undergo a spin reorientation transition which tips up the magnetization axis towards the hexagonal axis $\vec{c}$ \cite{Tie1991,Kuzmin2007}. This transition in NdCo$_5$ has recently attracted renewed attention due to a large associated rotating magnetocaloric effect \cite{Nikitin2010,Wang2019}. 

The RECo$_5$ family has been extensively studied experimentally for over 50 years. In particular, besides macroscopic magnetization measurements using magnetometers, measurements of microscopic magnetization distribution by polarized-neutron scattering (PNS) \cite{Boucherle1982} were carried out on single crystals for SmCo$_5$\cite{Givord1979} and NdCo$_5$\cite{Alameda1982}. These measurements allow unambiguously separating out the RE and TM contributions to the magnetization. Alameda {\it et al.} \cite{Alameda1982} thus found out that in  NdCo$_5$ the Nd GS moment is reduced by about 20\% compared to the saturation value of 3.27 $\mu_B$. This was puzzling since a full saturation was expected at low temperatures as confirmed by explicit calculations carried out with values within acceptable ranges for $B_{\mathrm{ex}}$ and the  "20" zonal low-rank $A_2^0\langle r^2 \rangle$ CFP \cite{Alameda1982}. The reduced Nd moment observed by Ref.~\onlinecite{Alameda1982} remains unexplained for almost 40 years, with previously reported CF schemes not able to account for it (see Table~\ref{table:CFPSs_comp}). 

Recently Delange {\it et al.}~\cite{Delange2017}  introduced a new approach to evaluating the CF. This methodology is based on the   DFT+dynamical mean-field theory (DFT+DMFT) in conjunction with the simple quasi-atomic Hubbard-I \cite{hubbard_1} treatment of RE 4$f$ shells and  employs an averaging scheme to remove the unphysical contribution\cite{Brooks1997} of DFT self-interaction error into the CF.  Delange {\it et al.}   successfully applied this methodology to SmCo$_5$ quantitatively reproducing the Sm 4$f$ CF GS measured by the PNS\cite{Givord1979} as well as the overall CF splitting in this intermetallic. 

In the present work we apply this method to determine the CFPs and $B_{\mathrm{ex}}$ in two easy-plane RECo$_5$ compounds, NdCo$_5$ and TbCo$_5$, evaluating their GS 4$f$ magnetic moments as well as RE single-ion anisotropy constants and their temperature dependence. Our crucial finding is that the sectoral "66" high-rank $A_6^6\langle r^6 \rangle$ CFP, often neglected in previous analyses, takes exceptionally large values in RECo$_5$. In NdCo$_5$ this CFP is shown  to freeze the GS magnetic moment below its fully saturated value thus explaining the result of Alameda {\it et al.}~\cite{Alameda1982}. The same CFP strongly enhances the easy plane magnetic anisotropy of NdCo$_5$, contradicting the erroneous belief according to which a "66" CFP would influence solely the polar magnetic anisotropy but not the energy difference between easy axis and easy plane. Even at elevated temperatures the easy plane anisotropy of NdCo$_5$ is significantly enhanced by the "66" CFP. This behavior is unexpected within the standard single-multiplet  framework (see, e. g., Ref.~\onlinecite{Kuzmin2007} for a review) and shown to stem from  $J$-mixing effects. Our resulting  anisotropy constants for NdCo$_5$ and their temperature dependence are in excellent agreement with experiment. Our analysis shows that the large "66" CFP originates in the hybridization mixing between 4$f$ and conduction states. It is expected to  be rather universal along the RECo$_5$  series. This is confirmed with TbCo$_5$, for which we also obtain a large value of "66" CFP though significantly reduced compared to NdCo$_5$. However, the impact of this "66" CFP on the TbCo$_5$ GS magnetism and magnetic anisotropy is found to be very weak, suggesting that this impact is element-sensitive. 

The paper is organized as follows: in Sec.~\ref{sec:method} we review the methodology used for the electronic structure calculations, establish the notation for the 4$f$ single-ion Hamiltonian and crystal-field parameters and recall the method for computing from ionic states the RE contribution to the magnetization distribution as probed by PNS. Our results are presented in Sec.~\ref{sec:results}, first on NdCo$_5$ then, more briefly, on TbCo$_5$. The origin of the large "66" CFP in RECo$_5$ is analyzed in Sec.~\ref{sec:analysis}. We list the calculated RE CF 4$f$ wave functions and CFPs for NdCo$_5$ and TbCo$_5$ in Appendix. 

\section{Method}\label{sec:method}

\subsection{Electronic structure and crystal field calculations}\label{subsec:elstruct}

For electronic structure calculations of the RECo$_5$ intermetallics we employed the self-consistent in charge density DFT+DMFT method of Refs.~\onlinecite{Aichhorn2009,Aichhorn2011}. It combines a full-potential linearized augmented planewave (FP-LAPW) band structure approach
\cite{Wien2k} and the DMFT implementation provided by the library "TRIQS" \cite{Parcollet2015,Aichhorn2016}. 

Calculations were carried out using the experimental hexagonal structure isotypic of CaCu$_5$ belonging to the space group $P6/mmm$, with the lattice parameters $a=$~5.00~\AA, $c=$~3.98~\AA\  for NdCo$_5$ and $a=$~4.95~\AA, $c=$~3.98~\AA\  for TbCo$_5$, and for the magnetically-ordered phase. We employed the local-spin density approximation to described the ordered Co magnetism.  The spin-orbit coupling was included within the standard second-variation procedure as implemented in Ref.~\onlinecite{Wien2k}, which is expected to be sufficient for the valence electronic states of RE ions. The RE 4$f$ shell was described within DMFT using the quasi-atomic Hubbard-I \cite{hubbard_1} approximation for the DMFT quantum impurity problem. Hereafter our {\it ab initio} appoach is  abbreviated as DFT+HubI. 

Wannier orbitals $\omega_{m\sigma}$ representing RE 4$f$ states (where $m$ and $\sigma$ are magnetic and spin quantum numbers, respectively) were constructed from the Kohn-Sham (KS) bands enclosed in a chosen energy window $\mathcal{W}$; this window must enclose at least  4$f$-like bands. In NdCo$_5$, similarly to previously studied\cite{Delange2017} SmCo$_5$ and light-RE Fe "1-12" systems, the RE 4$f$ bands are pinned at the KS Fermi level $E_{EF}^{KS}$, and we thus employed, unless noted otherwise, the same choice, $\mathcal{W}_s=[-2:2]$~eV relative to $E_{EF}^{KS}$, as in Ref~\onlinecite{Delange2017} . Test calculations using yet more narrow energy window ( $[-1:1]$~eV) produced similar results to those obtained with  $\mathcal{W}_s$. In contrast, with a wide-range energy window including all valence bands the RE 4$f$ ground state and CFPs are drastically modified, owing to the fact that the hybridization contribution to CFPs is in this case neglected by DFT+HubI, see the discussion in Sec.~\ref{sec:analysis} on the choice of RE 4$f$ orbitals in DFT+HubI calculations. In the case of TbCo$_5$ the 4$f$ KS bands shift significantly below the KS Fermi level in the course of DFT+HubI self-consistent calculations. Therefore, in that case we employed the same window range of 4~eV, but centered at the center-weight of the KS 4$f$ band, see Sec.~\ref{sec:analysis}. 


Within the Hubbard-I approximation the DMFT impurity problem is reduced\cite{Lichtenstein_LDApp} to diagonalization of the  Hamiltonian for a single 4$f$ shell:
\begin{equation}\label{eq:H_at_full}
 \hat{H}_{at}=\hat{H}_{1el}+\hat{H}_U=\sum_{mm'\sigma\sigma'}\epsilon_{mm'}^{\sigma\sigma'}f^{\dagger}_{m\sigma}, f_{m'\sigma'}+\hat{H}_U, 
 \end{equation}
 where $f_{m\sigma}$ ($f^{\dagger}_{m\sigma}$) is the creation (annihilation) operator for the RE 4$f$ orbital $m\sigma$ and $\hat{H}_U$ is the on-site Coulomb repulsion. The one-electron level-position matrix $\hat{\epsilon}$ reads\cite{Pourovskii2007}:
\begin{equation}\label{H1el_HI}
	\hat{\epsilon}=-\mu+\langle \hat{H}_{KS} \rangle^{ff} -\Sigma_{\mathrm{DC}},
\end{equation}
where $\mu$ is the chemical potential, $\langle \hat{H}_{KS} \rangle^{ff}=\sum_{\vk \in BZ}\hat{P}_{\vk} H_{KS}^{\vk}\hat{P}_{\vk}^{\dagger}$ is the Kohn-Sham Hamiltonian projected to the basis of 4$f$ Wannier orbitals $\omega_{m\sigma}$ and summed over the Brillouin zone, $\hat{P}_{\vk}$ is the corresponding projector between the KS and Wannier spaces \cite{Aichhorn2009,Aichhorn2016}, $\Sigma_{\mathrm{DC}}$ is the double counting correction term. 

The on-site Coulomb repulsion vertex $\hat{H}_U$ is specified for an $f$ shell by the Slater parameters F$^0$, F$^2$, F$^4$, F$^6$. Under the usual approximation of fixing the ratios F$^2$/F$^4$ and F$^2$/F$^6$  to the values obtained  experimentally\cite{Carnall1989} or in Hartree-Fock calculations for the corresponding free ions \cite{Freeman1962}, the vertex is determined by the two parameters, $U=F^0$ and the Hund's rule coupling $J_H$.  We employed F$^2$/F$^4=$ 1.5 and F$^2$/F$^6=$ 2.02. The  values  of 6.0  and 7.0~eV were used for the parameter $U$ of Nd and Tb, respectively, to take into account its expected increase along the RE series. We employed $J_H=$ 0.85~eV for Nd, in agreement with Ref.~\onlinecite{Delange2017} , the value  0.95~eV for $J_H$ of Tb was chosen in accordance with Ref.~\onlinecite{Carnall1989}. CFPs calculated with our approach have been shown\cite{Delange2017} to be weakly dependent on both $U$ and $J_H$.

Our self-consistent DFT+HubI calculations were carried out employing the self-interaction-suppressed scheme of Ref.~\onlinecite{Delange2017}. Namely, we averaged the Boltzmann weights of the eigenstates of $\hat{H}_{at}$ belonging to the atomic GS multiplet ($^4I_{9/2}$ and $^7F_6$ for  Nd and Tb, respectively). With all atomic states within the ground-state multiplet having the same occupancy\footnote{The Boltzmann weights of other states are negligible under the condition of temperature $T$ being much smaller than the intermultiplet splitting; this condition is satisfied for the 4$f$ shells of Nd and Tb for temperatures in the relevant range of several hundreds Kelvins.} one obtains a spherically-symmetric 4$f$ shell, similarly to a free RE atom. This procedure eliminates the unphysical contribution of the LDA self-interaction (SI) error to the CF splitting, since the SI contribution to $\hat{\epsilon}$  becomes orbitally independent in the case of a spherically-symmetric charge density. The same procedure also removes the spin polarization of the 4$f$ shell and, hence, its contribution to the LSDA exchange-correlation potential. The exchange field $B_{\mathrm{ex}}$ on the 4$f$ shell is in this case solely due to the magnetization density of Co sublattice. We thus neglect the contribution to $B_{\mathrm{ex}}$ due to the 4$f$-4$f$ inter-site exchange; this contribution, as mentioned in the introduction, is expected to be small in RECo$_5$ compounds.  The double-counting correction $\Sigma_{\mathrm{DC}}$ was hence calculated in the non-spin-polarized fully-localized limit \cite{Czyzyk1994} using the atomic occupancies~ \cite{Pourovskii2007} of the Nd or Tb 4$f$ shell.  

The CFPs are extracted from the converged one-electron level-position matrix $\hat\epsilon$ by fitting it to the form expected for the  corresponding RE ion embedded in a given crystalline environment:
\begin{equation}\label{eq:H_at}
\hat{\epsilon}=\hat{E}_0+\lambda \sum_{i}\hat{s}_i \hat{l}_i  +\hat{H}_{\mathrm{ex}}+\hat{H}_{\mathrm{ext}}+\hat{H}_{\mathrm{cf}},
\end{equation}
where the terms on the RHS stand successively for the uniform shift, the spin orbit coupling, the TM-RE exchange coupling, the Zeeman coupling $\hat{H}_{\mathrm{ext}}=-\mu_0\mathbf{H}_{\mathrm{ext}}\cdot\mathbf{M}$ of the RE moment $\mathbf{M}$ with an externally applied magnetic field $\mathbf{H}_{\mathrm{ext}}$ and the CF one-electron Hamiltonian. The TM-RE exchange coupling reads
$$
\hat{H}_{\mathrm{ex}}=2\mu_B B_{\mathrm{ex}}\mathbf{n}\cdot\mathbf{\hat{S}}_f,
$$
where the value of $B_{\mathrm{ex}}$  acting on the RE 4$f$-shell spin $\mathbf{\hat{S}}_f$ is determined by the RE-TM exchange coupling strength and the  TM-sublattice magnetization, which is directed along $\mathbf{n}$.

The RE site in the RECo$_5$ crystal structure has the point-group symmetry $6/mmm$, for which the CF contribution $\hat{H}_{\mathrm{cf}}$ to the one-electron level positions (\ref{eq:H_at}) reads
\begin{equation}\label{eq:H_cf}
\hat{H}_{\mathrm{cf}}= L_2^0 \hat{T}_2^0+ L_4^0 \hat{T}_4^0+ L_6^0 \hat{T}_6^0+ L_6^6 \hat{T}_6^6,
\end{equation}
by selecting as principal axis the hexagonal axis $\vec{c}$ ([001]), which is then the quantization axis of the 4$f$ electronic states. The $\hat{T}_k^q$ are the Hermitian Wybourne's tensor operators, related to the standard Wybourne's spherical tensor operators \cite{Wybourne1965} $\hat{C}_k^q$ as $\hat{T}_k^0 = \hat{C}_k^0$ and $ \hat{T}_k^{\pm|q|} = \sqrt{\pm1} \left[ \hat{C}_k^{-|q|}  \pm (-1)^{|q|} \hat{C}_k^{|q|} \right]$. The $L_k^q$ are the CFPs in the Weybourne's  convention.  

The CF Hamiltonian of RECo$_5$ intermetallics in the literature is often presented in the popular Stevens form:
\begin{eqnarray}\label{eq:H_cf_St}
\hat{H}^{St}_{\mathrm{cf}}= & \alpha_J A_2^0\langle r^2 \rangle  \hat{O}_2^0+\beta_J A_4^0\langle r^4 \rangle  \hat{O}_4^0 \\
& + \gamma_J \left[A_6^0\langle r^6 \rangle \hat{O}_6^0+ A_6^6\langle r^6 \rangle \hat{O}_6^6\right] , \nonumber
\end{eqnarray}
where the $\hat{O}_k^q$ are the Stevens operators\cite{Stevens1952} acting on many-electron 4$f$ wavefunctions within the atomic GS multiplet, for example
\begin{equation}\label{eq:Os}
\hat{O}_2^0=3\hat{J}_z-J(J+1),   \textrm{     }
\hat{O}_6^6=\frac{1}{2}(\hat{J}_+^6+\hat{J}_-^6), \cdots
\end{equation}
 $\alpha_J$, $\beta_J$, and $\gamma_J$ are the Stevens factors $\Theta_k$ for $k=$~2, 4, and 6, respectively, for a given value of the total angular momentum $J$. $A_k^q\langle r^q \rangle$ are the CFPs in the Stevens convention, related to the Wybourne notation by $A_k^q\langle r^q \rangle=\lambda_{kq} L_k^q $, with the prefactors $\lambda_{kq}$ tabulated elsewhere.\cite{Newman2000,Mulak2000} We shall use the Stevens convention for our calculated CFPs to ease comparison with the literature. 

The self-consistent DFT+HubI calculations were converged to less than 1\% with respect to the values of CFPs, which were obtained by fitting of {\it ab initio} level positions $\hat{\epsilon}$ to the form (\ref{eq:H_at}). We also performed calculations with the CF description suited to the choice of the binary axis $\vec{a}$ ([100]) as principal axis. In this setting the unit cell is orthorhombic with the lattice parameters $c$, $\sqrt{3}a$ and $a$ in terms of the original hexagonal cell parameters. All $A_k^q\langle r^k \rangle$ for even  positive $q \le k$ are nonzero in this setting. The resulting CFPs  of the orthorhombic cell were found to agree with those of the hexagonal cell after the rotation by Euler angle $\beta=\pi/2$. 

Once the CFPs are obtained from converged DFT+HubI calculations we extract RE magnetic anisotropy by solving the full-shell Hamiltonian (\ref{eq:H_at_full}) at various orientation $\mathbf{n}$ of the exchange field $B_{\mathrm{ex}}$, with the level positions given by eq.~\ref{eq:H_at} and $H_{\mathrm{cf}}$ by eq.~\ref{eq:H_cf}. All inter-multiplet mixing effects are thus included in these calculations. For the sake of comparison and when it is noted explicitly we perform also single GS multiplet (GSM) calculations using the Stevens operator form (\ref{eq:H_cf_St}) and diagonalizing the corresponding Hamiltonian  $\hat{H}^{St}_{\mathrm{cf}}+\hat{H}_{\mathrm{ex}}$ defined in the GSM space. The $B_{\mathrm{ex}}$ term in this space is written 
\begin{equation}\label{H_ex_GSM}
\hat{H}_{\mathrm{ex}}=\Delta_{\mathrm{ex}}\mathbf{n}\mathbf{\hat{J}}; \textrm{  } \Delta_{\mathrm{ex}}=2 (g_J-1)\mu_BB_{\mathrm{ex}},
\end{equation}
where $g_J$ is the gyromagnetic ratio for the GSM.

\subsection{Calculations of magnetization distribution} 

RE contribution to magnetization distribution $\vec{\mathcal{M}}(\vec{r})$ as probed by PNS can be inferred from ionic states underlying the fit of {\it ab initio} matrix $\hat{\epsilon}$ to the form (\ref{eq:H_at}). $\vec{\mathcal{M}}(\vec{r})$ is experimentally generated from neutron magnetic structure factors $\vec{\mathcal{F}}^\bot (\vec{\varkappa}) =  \{\vec{\varkappa} \wedge \int \vec{\mathcal{M}} (\vec{r}) e^{i \vec{\varkappa}\cdot\vec{r}} d\vec{r} \wedge \vec{\varkappa}\} / ({\vec{\varkappa}\cdot\vec{\varkappa}})$, which in centrosymmetric collinear ferrimagnets are precisely determined by collecting the intensity ratios of diffracted neutrons on all accessible reciprocal lattice vectors $\vec{\varkappa}$ for ingoing neutrons polarized parallel and antiparallel to magnetization \cite{Boucherle1982}. Generally, the most accessible reciprocal lattice vectors $\vec{\varkappa}$ are those lying in the plane perpendicular to magnetization for which $\vec{\mathcal{F}}^\bot (\vec{\varkappa})$ is parallel to magnetization. The amplitude $\mathcal{F}^\bot (\vec{\varkappa})$ of $\vec{\mathcal{F}}^\bot (\vec{\varkappa})$ is then interpreted as a Fourier coefficient of the amplitude $\mathcal{M}(\vec{r})$ of the projection of $\vec{\mathcal{M}}(\vec{r})$ on the plane perpendicular to $\vec{\mathcal{M}}(\vec{r})$. The RE part of $\vec{\mathcal{F}}^\bot (\vec{\varkappa})$ can be evaluated over its electronic spectrum as $\vec{\mathcal{F}}_{RE}^\bot (\vec{\varkappa}) = \langle \int \{ - \vec{\varkappa} \wedge \vec{\nabla}_{\vec{r}} + \vec{\varkappa} \wedge \hat{\vec{s}}(\vec{r}) \wedge \vec{\varkappa} \} e^{i \vec{\varkappa} \cdot\vec{r}} d\vec{r}/ ({\vec{\varkappa}\cdot\vec{\varkappa}}) \rangle_{RE} W_{RE}(\vec{\varkappa}) = \vec{\mathcal{E}}_{RE} (\vec{\varkappa}) W_{RE}(\vec{\varkappa})$ where the expression inside the curly brackets distinguishes orbital and spin contributions and $W_{RE}$ stands for the RE Debye-Waller vibrating factor. $\langle \cdots \rangle_{RE}$ symbolizes quantum statistical average. At low temperatures it reduces to a matrix element over the ground state $\Psi_{GS}^{RE}$. Using the tensor-operator formalism, \cite{Lovesey1969} the spherical components of the vibrating-free neutron magnetic structure factor $\vec{\mathcal{E}}_{RE} (\vec{\varkappa})$ can be written, in units of Bohr magneton ($\mu_B$), in the form
 \begin{eqnarray}
 \label{eq:msf}
{\mathcal{E}_{RE}^M (\vec{\varkappa})}_q^1 & = & -4\sqrt{\pi} \sum_{K,Q} Y_Q^K(\theta_{\vec{\varkappa}},\phi_{\vec{\varkappa}})  \sum_{K^\prime,Q^\prime} \langle K Q K^\prime Q^\prime | 1 q \rangle 
\nonumber \\
& \times  & \bigg\{ \sum_{\substack{\theta J M \\ \theta^\prime J^\prime M^\prime}} \langle \theta^\prime J^\prime M^\prime | \Psi_{GS}^{RE}\rangle  \langle \Psi_{GS}^{RE} | \theta J M \rangle
\nonumber \\
&   &  \times (\mathfrak{A}_{KK^\prime} + \mathfrak{B}_{KK^\prime}) \langle K^\prime Q^\prime J^\prime M^\prime | J M \rangle \bigg\}
\end{eqnarray}
%
%
using the basis of $4f$ ionic states $|\theta J M \rangle \equiv |4f^n \upsilon LS J M \rangle$ with total orbital momentum $L$, total spin $S$ and total angular momentum $J$ with azimuthal component $M$. The $Y_Q^K ~ (-K \leq Q \leq K)$ stand for spherical harmonics of order $K$. $(\theta_{\vec{\varkappa}},\phi_{\vec{\varkappa}})$ are the azimuthal and polar angles of $\vec{\varkappa}$. $\langle \cdot \cdot \cdot \cdot | \cdot\cdot \rangle$  symbolizes  Clebsh-Gordon coefficients. $\mathfrak{A}_{KK^\prime}$ and $\mathfrak{B}_{KK^\prime}$ arise respectively from the neutron scattering on the orbital part and on the spin part of the electronic wavefunction. They depend on the radial part $R_{4f}$ of this wavefunction through the radial integrals $\langle j_K (\varkappa) \rangle = \int_0^\infty dr~r^2 | R_{4f}(r) |^2 j_K(\varkappa r)$, 
where $ j_K$ is the spherical Bessel function of order $K$. These were numerically calculated from the relativistic Dirac-Fock Hamiltonian for all the trivalent RE ions. \cite{Freeman1979} The tabulated values were approximated by analytic functions. \cite{Brown2006} The explicit formula of $\mathfrak{A}_{KK^\prime}$ and $\mathfrak{B}_{KK^\prime}$ are detailed in Ref. \onlinecite{Lovesey1969} and involve, besides $nj-$symbols, parent states and coefficient of fractional parentage that can be found {\it e.g.} in Ref. \onlinecite{Nielson1963}. Note that it may be inferred from properties of $nj-$symbols that $\mathfrak{A}_{KK^\prime}$ is null unless $K$ is even, $K^\prime$ is odd and $K=K^\prime\pm1$. Moreover $\mathfrak{A}_{K^\prime+1K^\prime} = \{K^\prime/(K^\prime+1)\}^\frac{1}{2} ~ \mathfrak{A}_{K^\prime-1K^\prime}$. For $f$ states, $K^\prime = 1, 3, 5$. It may also be inferred that $\mathfrak{B}_{KK^\prime}$ is null unless $K$ is even, $K^\prime$ is even and $K=K^\prime$ or $K$ is even, $K^\prime$ is odd and $K=K^\prime\pm1$ in which case $\mathfrak{B}_{K^\prime+1K^\prime} = \{K^\prime/(K^\prime+1)\}^\frac{1}{2} ~ \mathfrak{B}_{K^\prime-1K^\prime}$. For $f$ states, $K^\prime = 2, 4, 6$ for $K = K^\prime$ and $K^\prime =1, 3, 5, 7$ for $K=K^\prime\pm1$.
%
%

\section{Results} \label{sec:results}

\subsection{4$f$ ground state and  zero-temperature magnetization in NdCo$_5$}

The converged GS of Nd 4$f^3$ shell obtained by the self-interaction suppressed DFT+HubI calculations in NdCo$_5$ reads
\begin{eqnarray}
 \label{eq:Nd_GS}
\Psi_{GS}^{Nd} & = & 0.827|9/2 -9/2\rangle-0.536|9/2 -5/2\rangle
\nonumber\\
&  & -0.089|9/2 -1/2\rangle
\nonumber\\
&  & -0.096|11/2 -9/2\rangle+0.094|11/2 -5/2\rangle
\end{eqnarray}
where $|J M\rangle$ is a shorthand notation for the basis states $|4f^3 \upsilon ~ L=6 ~ S=3/2~ J M \rangle$ and the quantization axis is chosen along the binary axis $\vec{a}$ ([100]) of the hexagonal structure, i.e. along the GS magnetization direction\cite{Alameda1982,Bartashevich1993,Zhang1994} in NdCo$_5$. 
   \begin{figure}[b]
   	\begin{center}
   		\includegraphics[width=0.90\columnwidth]{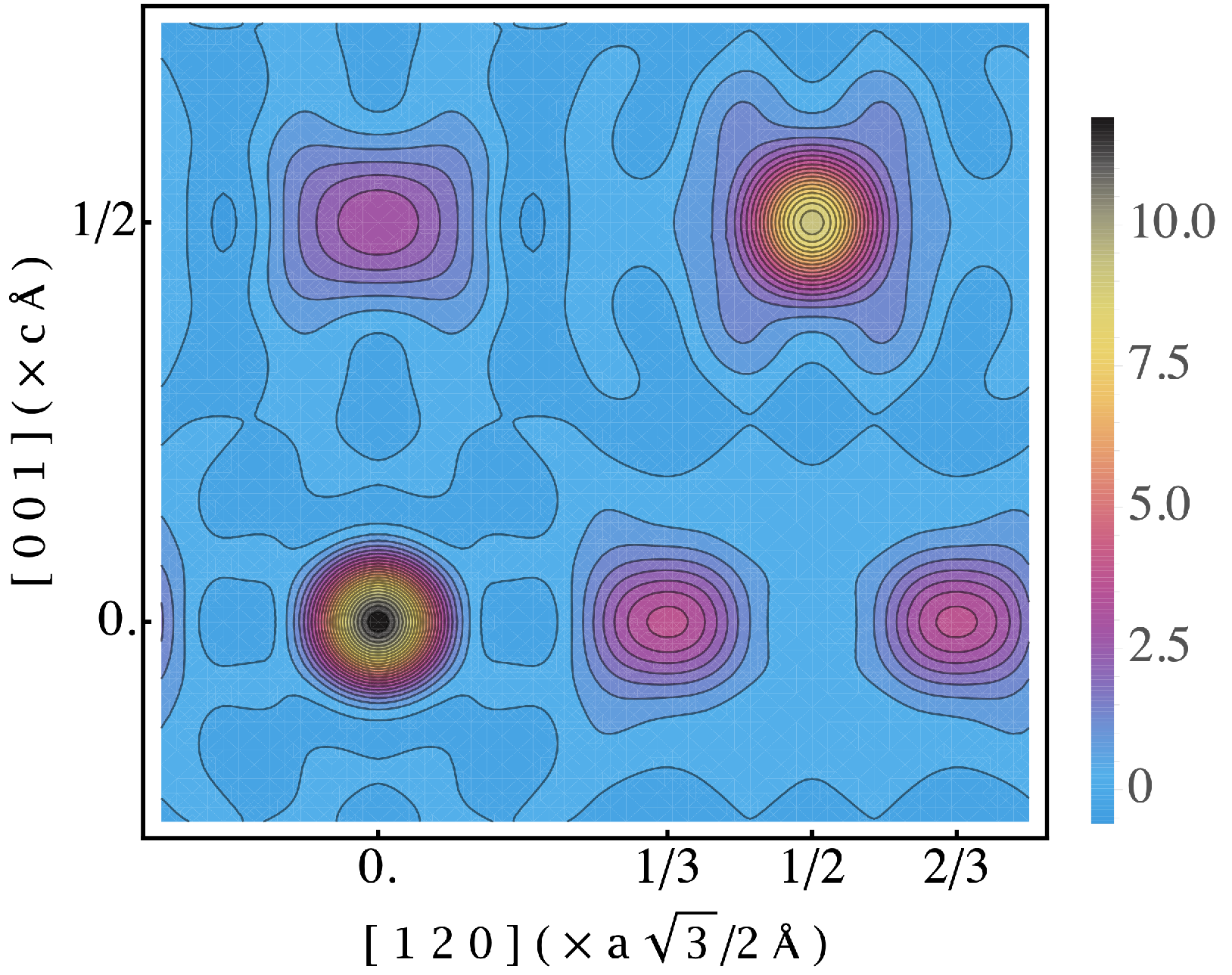}
   	\end{center}
   	\caption{Magnetization distribution $\mathcal{M}(\vec{r})$ in NdCo$_5$ as projected in the plane ($\vec{c}$ ([001]), $\vec{c}\wedge\vec{a}$ ([120])) perpendicular to the orientation $\vec{a}$ ([100]) of $\vec{\mathcal{M}}(\vec{r})$, inferred through Fourier summation from neutron magnetic structure factors reported in Ref.~\onlinecite{Alameda1982}. The Nd ion on site $1a$ is projected at position $(0,0)$, the Co ions on site $2c$ are projected at positions $(0,1/3)$ and $(0,2/3)$ and the Co ions on site $3g$ are projected at positions $(0,1/2)$ and, for two of them, $(1/2,1/2)$. The Nd contribution to this experimental magnetization distribution map in projection is thus fully separated from the Co contributions.}
   	\label{fig:MD_NdCo5}
   \end{figure}
Table~\ref{table:WFs_NdCo5} in Appendix provides the complete list of Nd CF eigenstates. The first excited state is 220~K above in energy, hence, the low-temperature Nd magnetization is determined by the GS $\Phi_{GS}^{Nd}$ and equal to 2.66~$\mu_B$, which is significantly lower than the saturated value of 3.27~$\mu_B$  of the GS $^4I_{9/2}$ multiplet of Nd$^{3+}$. Indeed, the GS wavefunction (\ref{eq:Nd_GS}) features a large contribution from the component $|9/2 -5/2\rangle$ besides the dominating component $|9/2 -9/2\rangle$. The unsaturation of the Nd magnetic moment in NdCo$_5$ had been previously evidenced by Alameda {\it et al.}~\cite{Alameda1982} following a PNS experiment. The measured magnetic structure factors they provide, all at reciprocal lattice vectors $\vec{\varkappa}$ perpendicular to magnetization, allow generating, through Fourier summation, the magnetization distribution $\mathcal{M}(\vec{r})$ as projected on the plane ($\vec{c}$~([001]), $\vec{c}\wedge\vec{a}$~([120])) perpendicular to $\vec{a}$ ([100]). As displayed in Fig.~\ref{fig:MD_NdCo5}, it exhibits little if any overlapp between Nd contribution and Co ones. Integrating this experimental magnetization distribution over ovoid and rectangular surfaces of increasing size centred on the Nd crystallographic site leads to a magnetic moment that never exceed 2.70~$\mu_B$ except when the surfaces start overlapping the magnetization distribution visually ascribable to Co. However, this maximum might not correspond to the true Nd magnetic moment since not all the magnetic structure factors were measured. 

   \begin{figure}[t]
   	\begin{center}
   		\includegraphics[width=0.80\columnwidth]{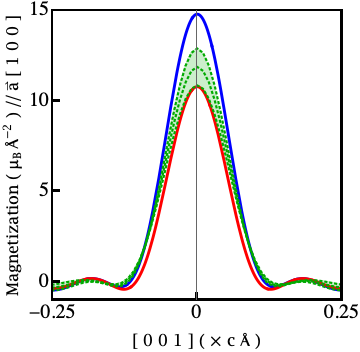}
   	\end{center}
   	\caption{Magnetization distribution $\mathcal{M}(0, z)$ in NdCo$_5$ along a segment crossing Nd position parallelly to the axis $\vec{c}$ ([001]. The experimental profile is in dotted green. It is plotted with confidence bands inferred from experimental uncertainties reported in Ref.~\onlinecite{Alameda1982} for the measured magnetic structure factors. The curve in blue stands for the profile calculated from the wavefunction $\Phi_{SAT}^{Nd} = |9/2 -9/2\rangle$. The  curve in red stands for the profile calculated from the wavefunction $\Psi_{GS}^{Nd}$ (see \ref{eq:Nd_GS})}
   	\label{fig:MDpc_NdCo5}
   \end{figure}
The experimental magnetization distribution can be relevantly compared to the one inferred from magnetic structure factors obtained from a given $4f$ wavefunction through eq.~\ref{eq:msf} provided that the calculations are performed on the same reciprocal lattice vectors as in the experiment. This is illustrated in Fig.~\ref{fig:MDpc_NdCo5} by the magnetization distribution obtained from wavefunctions $\Psi_{SAT}^{Nd} = |9/2 -9/2\rangle$ and $\Psi_{GS}^{Nd}$, corresponding to  saturated and unsaturated Nd magnetic moments, respectively. The magnetization distribution computed from $\Psi_{SAT}^{Nd}$ is clearly larger than the experimental one, beyond experimental uncertainties. On the other hand, the magnetization distribution computed from $\Psi_{GS}^{Nd} $ is in agreement with experiment, inside experimental confidence bands.

Alameda {\it et al.} analyzed their data by means of a parametric modelling for the measured magnetic structure factors $\mathcal{F}^\bot (\vec{\varkappa})_0^1$. The Nd contribution $\mathcal{F}_{Nd}^\bot (\vec{\varkappa})_0^1$ was computed assuming a GS wavefunction in the form
\begin{equation}\label{eq:Nd_GS_exp}
 \Psi_{REF}^{Nd}=\alpha|9/2-9/2\rangle \pm \sqrt{1-\alpha^2}|9/2-5/2\rangle
 \end{equation}
thus neglecting the contribution of excited multiplets. The quantities $\mathfrak{A}_{KK^\prime}$ and $\mathfrak{B}_{KK^\prime}$ in eq.~\ref{eq:msf} can in that case be readily evaluated using tabulated coefficients \cite{Lander1970}. The Co contribution $\mathcal{F}_{Co}^\bot (\vec{\varkappa})_0^1 = \sum_i \mathcal{E}_{Co}^i (\vec{\varkappa})_0^1~ e^{i \vec{\varkappa} \cdot \vec{r}_i}~W_{Co}^i(\vec{\varkappa})$, where $\vec{r}_i$ defines the position of the $i^{th}$ Co atom in the unit cell and $W_{Co}^i(\vec{\varkappa})$ its Debye-Waller vibrating factor, was evaluated according to the same approach as in a previous work on YCo$_5$ \cite{Schweizer_1980}. In result, a factor $\alpha=$ 0.83, determining the relative weight of $|9/2 -9/2\rangle$ and $|9/2 -5/2\rangle$ in the GS, was obtained in Ref.~\onlinecite{Alameda1982} by fitting the model to reproduce the measured  magnetic structure factors. The GS moment of Nd, calculated from eq.~\ref{eq:Nd_GS_exp} with this value of $\alpha$, is 2.82 $\mu_B$. We obtain 2.84~$\mu_B$ by applying the same procedure to (\ref{eq:Nd_GS}), i.e. by neglecting the contributions of excited multiplets thus normalizing the GS wave function to 1 within the GS multiplet. 
   \begin{figure}[b]
   	\begin{center}
   		\includegraphics[width=0.80\columnwidth]{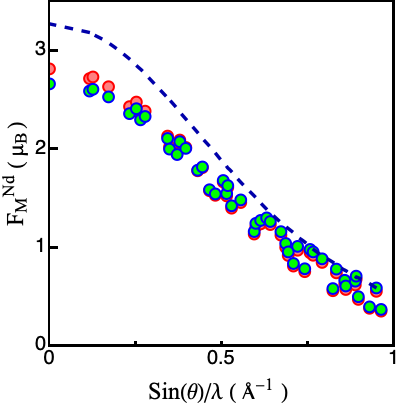}
   	\end{center}
   	\caption{Nd magnetic structure factors in NdCo$_5$. The red filled disks correspond to the experimental values refined in Ref.~\onlinecite{Alameda1982}. The cyan filled disks are the values computed from the full wavefunction $\Psi_{GS}^{Nd}$, eq.~\ref{eq:Nd_GS}. The dashed curve goes through the isotropic values computed from the saturated-state wavefunction $\Psi_{SAT}^{Nd} = |9/2 -9/2\rangle$.} 
   	\label{fig:MDSF_NdCo5}
   \end{figure}

The refined Nd contribution $(\mathcal{F}_{Nd}^\bot (\vec{\varkappa})_0^1)_{REF}$ to the magnetic structure factors obtained using $\alpha=$ 0.83 in eq.~\ref{eq:Nd_GS_exp} is displayed in Fig.~\ref{fig:MDSF_NdCo5}. As shown in Fig.~2 of Alameda {\it et al.} it coincides, within experimental error bars, with the experimentally measured structure factors of Nd. In Fig.~\ref{fig:MDSF_NdCo5} we  also show, for the same reciprocal lattice vectors $\vec{\varkappa}$, the magnetic structure factors $\mathcal{F}_{Nd}^\bot (\vec{\varkappa})_0^1$ computed using eq.~\ref{eq:msf}  from the wavefunction $\Psi_{GS}^{Nd}$,  eq.~\ref{eq:Nd_GS}. The structure factors $(\mathcal{F}_{Nd}^\bot (\vec{\varkappa})_0^1)_{SAT}$ computed from the fully saturated ground state $\Psi_{SAT}^{Nd} = |9/2 -9/2\rangle$ are also shown. The latter are isotropic, i.~e. they exhibit no dependence  on the direction of  $\vec{\varkappa}$, and thus $(\mathcal{F}_{Nd}^\bot (\vec{\varkappa})_0^1)_{SAT}$ collapse into a single line when plotted as a function of the  reciprocal lattice vector length $\varkappa = 4\pi \sin{(\theta)}/\lambda$.  $(\mathcal{F}_{Nd}^\bot (\vec{\varkappa})_0^1)_{SAT}$ is also clearly larger than both experimental $(\mathcal{F}_{Nd}^\bot (\vec{\varkappa})_0^1)_{REF}$ and our theoretical $\mathcal{F}_{Nd}^\bot (\vec{\varkappa})_0^1$ , especially at low reciprocal distance $\varkappa$. Theoretical $\mathcal{F}_{Nd}^\bot (\vec{\varkappa})_0^1$ is in an almost perfect agreement with $(\mathcal{F}_{Nd}^\bot (\vec{\varkappa})_0^1)_{REF}$ showing a similar anisotropy. The effect of the multiplet mixing is mostly manifest at low reciprocal distance $\varkappa$ where $\mathcal{F}_{Nd}^\bot (\vec{\varkappa})_0^1$ is noticeably lower than $(\mathcal{F}_{Nd}^\bot (\vec{\varkappa})_0^1)_{REF}$.

Alameda {\it et al.} found their result on the Nd magnetic moment puzzling, as large $B_{\mathrm{ex}}$ induced by the ferromagnetic Co sublattice in $RE$Co$_5$ was expected to saturate the $RE$ moment at low temperatures. Indeed, assuming a reasonable upper limit of the value of low-rank CFP $A_2^2\langle r^2 \rangle\approx$ -450~K and an equally reasonable value of $B_{\mathrm{ex}}\approx $ 300~T they obtained a fully-saturated GS with the magnetic moment of 3.27~$\mu_B$. However, in their analysis the higher-rank CFPs in (\ref{eq:H_cf_St}) were assumed to be irrelevant and were therefore neglected.
   \begin{figure}[hb]
   	\begin{center}
   		\includegraphics[width=0.90\columnwidth]{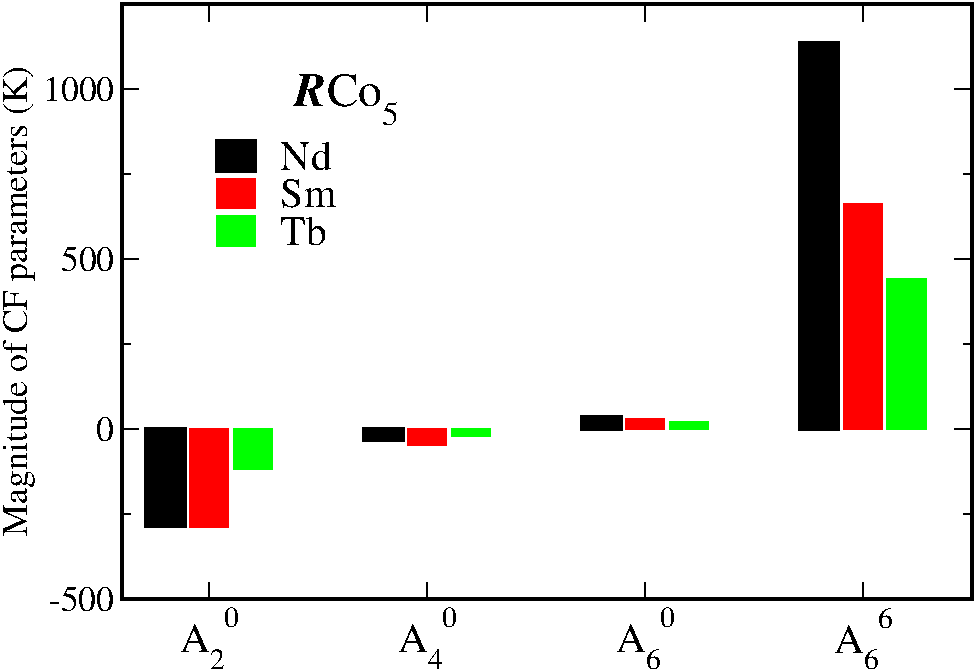}
   	\end{center}
   	\caption{Calculated  CFPs $A_k^q\langle r^q \rangle$ in RECo$_5$ ($R$=Nd,Sm, and Tb). "$\langle r^q \rangle$" is omitted from the axis labels for brevity. These CF parameters  are defined in a coordination frame with $z||c$ and $x||a$. Data for SmCo$_5$ are obtained from DFT+HubI calculations of Ref.~\onlinecite{Delange2017}; we assumed non-spin-polarized CFPs in fitting (eqs.~\ref{eq:H_at} and~\ref{eq:H_cf}) instead of spin-polarized ones as in Ref.~\onlinecite{Delange2017}. Notice the very large values of $A_6^6\langle r^6 \rangle$  in all three compounds.}
   	\label{fig:CF}
   \end{figure}
 
The CFPs  extracted from the converged DFT+HubI level positions (\ref{H1el_HI}) by fitting them to the form (\ref{eq:H_at}) are displayed in Fig.~\ref{fig:CF} . The fitted value of spin-orbit coupling $\lambda=$ 126~meV is in a good agreement with the experimental value of 110~meV for Nd$^{3+}$ impurity embedded into a crystalline host\cite{Carnall1989}. One may notice negative  $A_2^0\langle r^2 \rangle=-285$~K corresponding to an in-plane anisotropy experimentally observed in NdCo$_5$, but also a very large value for the calculated $A_6^6\langle r^6 \rangle$ ("66")  CFP, reaching 1134~K in NdCo$_5$.  

 	 \begin{table*}[hbt!]
 	 	 \onecolumngrid
 	 	\centering
 	 	\caption{\label{table:CFPSs_comp}  CFPs (in K) and exchange field $B_{\mathrm{ex}}$ (in Tesla) of NdCo$_5$ reported in previous theoretical and experimental works compared to the present one. The coefficient $\alpha$ in the wave function $\Psi_{REF}^{Nd}$, eq.~\ref{eq:Nd_GS_exp}, and corresponding GS magnetic moment (in $\mu_B$) calculated from given CFPs and $B_{\mathrm{ex}}$ are listed in the last two columns. {\it Ab initio} works are marked by ${^*}$. The measured $\alpha$ and the corresponding GS moment are given in the last line.} 
 	 	\begin{ruledtabular}
 	 	\begin{tabular}{l c c c c c c c}
 	 		&  $A_2^0\langle r^2 \rangle$ &  $A_4^0\langle r^4 \rangle$  & $A_6^0\langle r^6 \rangle$  &  $A_6^6\langle r^6 \rangle$  & $B_{\mathrm{ex}}$ & $\alpha $ &  $|M^{GSM}_{Nd}|$ \\			
 	 		\hline
 	 		Radwansky \onlinecite{Radwanski1986} & -210 & - & - & -& 151 & 1.0 & 3.26 \\
Zhao {\it et al.} \onlinecite{Tie1991} & -510 & 0 & 7 & 143 & 558 &  1.0 & 3.27 \\
Zhang {\it et al.}\onlinecite{Zhang1994}\footnote{Zhang {\it et al.}\cite{Zhang1994} report two sets of values for the CFPs and $B_{\mathrm{ex}}$}& -397 & -0.9 & 13.1 & 816 & 203 & 0.91 & 3.02 \\
& -482 & -0.9 & 13.1 & 816 & 393 & 0.97 & 3.19 \\
Novak$^*$ \onlinecite{Novak1996} \footnote{Novak\cite{Novak1996} does not report $B_{\mathrm{ex}}$, we thus employ two values representing the bounds of its generally accepted range}   & -288 & -44.7 & 11.3 & 573 & 150 & 0.87 & 2.93 \\   
& -288 & -44.7 & 11.3 & 573 & 450 &  0.96 & 3.18 \\                                             
Patrick \& Staunton$^*$ \onlinecite{Patrick2019}  & -415 & -26 & 5.4 & 146 & 252 & 1.0 & 3.27 \\
This work$^*$ & -285 & -33 & 36 & 1134 & 292 & 0.84 & 2.84 \\
Experiment~\onlinecite{Alameda1982} & & & & & & 0.83 & 2.82 \\                                                                                                    
 	 	\end{tabular}
 	 	\end{ruledtabular}
 	 \end{table*}
 	
In order to identify the impact of this large "66" CFP the CF level scheme was also calculated by setting it to zero. The resulting GS wave function is purely $|9/2-9/2\rangle$ corresponding to the fully saturated Nd moment. Hence, it is precisely this CFP that is preventing the full saturation of low-temperature Nd moment in NdCo$_5$.  

In Table~\ref{table:CFPSs_comp} we compare our calculated  CFPs and $B_{\mathrm{ex}}$ with experimental and theoretical values reported for NdCo$_5$  in the literature. The experimental values  in Table~\ref{table:CFPSs_comp} are obtained from fitting either to high-field magnetization curves or to the temperature dependence of magnetic anisotropy. The theoretical values are obtained by the DFT employing the open-core treatment for Nd 4$f$. In spite of the large discrepancies between different references one may notice that the "66" CFP values reported so far are significantly smaller than our calculated value, while our "20" CFP and $B_{\mathrm{ex}}$ are in the middle of literature values. For each set of CFPs$+B_{\mathrm{ex}}$ we compute the value of $\alpha$ as described above as well as the  Nd moment from the corresponding single-multiplet GS wave function (\ref{eq:Nd_GS_exp}).   One sees that none of previous CFP schemes, in spite of significant differences between them,  is able to account for the large admixture of $M=-5/2$ to the GS found by Alameda {\it et al.} and the corresponding reduction of the moment. The "freezing" of Nd GS moment thus represents a direct indication of the huge value of the "66" CFP. As we  argue in Sec.~\ref{sec:analysis} this value arises from the hybridization between localized 4$f$s and itinerant states, which is neglected within the "open-core "framework.

\begin{figure*}
	\begin{minipage}[b][6cm]{0.9\linewidth}
		\subfloat[ ]{\includegraphics[scale=0.33]{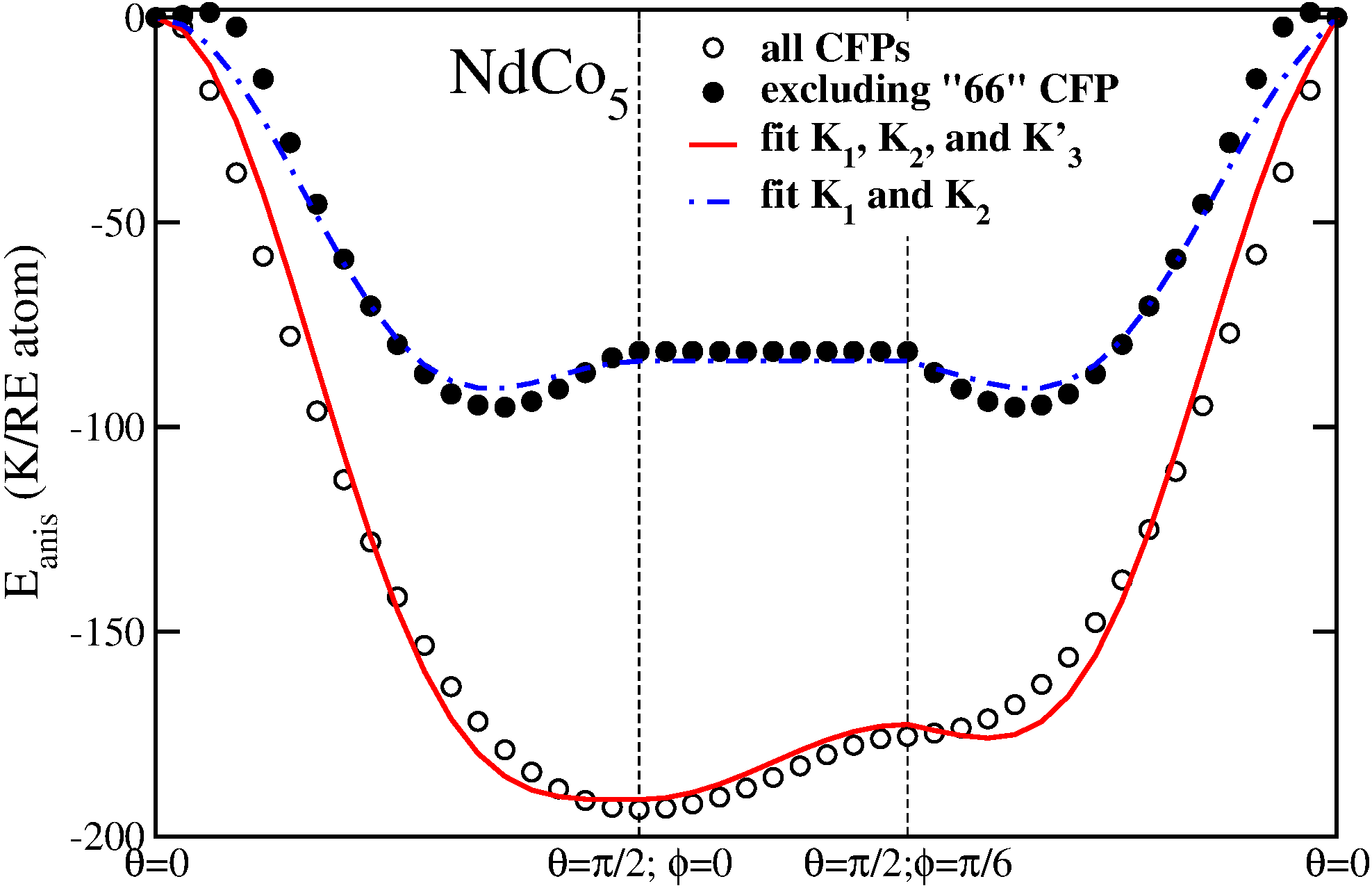}}
		\subfloat[ ]{\includegraphics[scale=0.33]{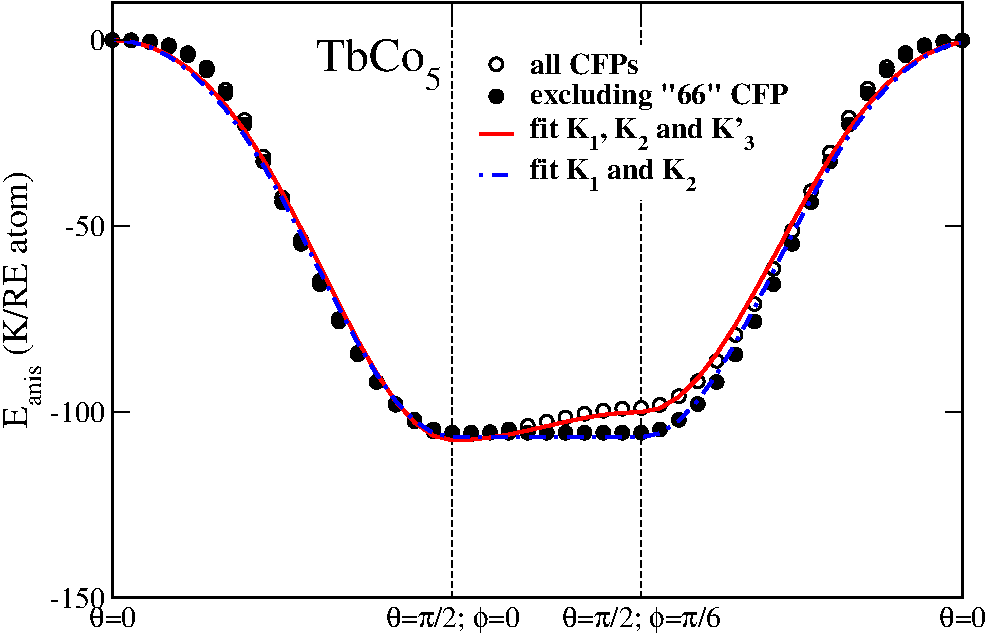}}
	\end{minipage}
	\caption{Ground state energy, $E_{GS}$, of rare-earth 4$f$ shell in (a) NdCo$_5$ and (b) TbCo$_5$ as a function of the exchange field's direction $\mathbf{n}$. The direction $\mathbf{n}$ is specified by the azimuthal angle $\theta$ and polar angle $\phi$. 
		Empty and filled circles indicate the  values computed by direct diagonalization of the Hamiltonian (\ref{H1el_HI})  constructed with and without the CF parameter $A_6^6\langle r^6 \rangle$, respectively. The lines are  a least-square fit of calculated $E_{GS}$  to the anisotropy-energy expression (\ref{eq:E_anis}) with the anisotropy constants specified in the legend. }
		\label{fig:K_fit}
	\end{figure*}


 \begin{table}[h]
 	\centering
 	\caption{\label{table:Ks} Zero-temperature RE single-ion anisotropy constants and magnetocrystalline anisotropy  energy (MAE), in units of K/f.u.. The values in parenthesis are obtained by the  Suscksmith-Thompson formula; other values are extracted by fitting the angular dependence of the calculated MAE (Fig.~\ref{fig:K_fit})  to eq. \ref{fig:K_fit}. For the anisotropy constant of Co sublattice, $K_1^{Co}$, we took the value of 45 K/(f.u.) measured in YCo$_5$.  Higher-order anisotropy constants of Co are negligible in accordance with experiment.\cite{Alameda1981} Experimental values (at $T=$ 4.2~K) from Refs.~\onlinecite{TATSUMOTO1971,Ermolenko1976} and  \onlinecite{Ermolenko1980} are indicated by superscripts $^a$, $^b$ and $^c$, respectively. }
 	\begin{ruledtabular}
 	\begin{tabular}{c  c  c  c}
 		 \multicolumn{4}{c}{NdCo$_5$}  \\
 		& with  $A_6^6\langle r^6 \rangle$ &  w/out $A_6^6\langle r^6 \rangle$  & Exp.  \\
 		\hline
 		$K_1$ & -393 & -231 & -510$^{c}$  \\
 		$K_1+K_1^{Co}$ & -348 (-211) & -186 &-244$^{a}$, -212$^{b}$,  -468$^{c}$ \\
 		$K_2$ & 211 (91)  & 147   &   119$^{a}$, 87$^{b}$, 193$^c$   \\
 		$K'_3$ & -9 & - & -  \\
 		MAE  & 	-148 (-120) & -37 & -125$^{a}$,  -125$^{b}$, -275$^c$   \\
 	\end{tabular}
 	\end{ruledtabular}
 	\vspace{3mm}
 	\begin{ruledtabular}
 	 	\begin{tabular}{c  c  c  c }
 	 		  \multicolumn{4}{c}{TbCo$_5$} \\
 	 		& with  $A_6^6\langle r^6 \rangle$ &  w/out $A_6^6\langle r^6 \rangle$  & Exp.  \\
 	 		\hline
 	 		$K_1$ & -59 & -64 & -99$^{c}$ \\
 	 		$K_1+K_1^{Co}$ &  -14 & -19 & -57$^{c}$ \\
 	 		$K_2$ &  -45 & -43 & -36 \\
 	 		$K'_3$ & -4 & -  & - \\
 	 		MAE  &  -63 & -62 & -93$^{c}$ \\
 	 	\end{tabular}
 	 \end{ruledtabular}
 \end{table}

\subsection{Zero-temperature magnetic anisotropy of NdCo$_5$}

Let us now analyze the impact of "66" CFP on the magnetocrystalline anisotropy energy (MAE). The MAE of a hexagonal crystal reads:
 \begin{eqnarray}\label{eq:E_anis}
 E_{anis}(\theta, \phi) & =  K_1 \sin^2 \theta + K_2 \sin^4 \theta+ K_3  \sin^6 \theta  \\
  &   + K_3'  \sin^6 \theta \cos 6\phi, \nonumber
 \end{eqnarray}
 where $\theta$ and $\phi$ are azimuthal and polar angles, respectively, of the magnetization direction in the reference frame with $z||c$ and $x||a$. The RE macroscopic anisotropy constants $K_i$  are determined by the interplay of $B_{\mathrm{ex}}$ and CFPs. In order to elucidate the impact of $A_6^6\langle r^6 \rangle$ on the Nd single-ion  anisotropy in NdCo$_5$ we numerically evaluated the Nd SIA constants $K_i$. To that end we diagonalized the Hamiltonian (\ref{eq:H_at}) parametrized by the calculated values of CFPs, $B_{\mathrm{ex}}$, and $\lambda$, varying the direction $\mathbf{n}$ of $B_{\mathrm{ex}}$ (i.~e. the direction of magnetization of the Co sublattice). We obtained a strong in-plane Nd single-ion anisotropy, with the easy direction along the $a$ direction of the hexagonal unit cell, as  seen from the calculated evolution of the GS energy along a chosen path in the $(\theta,\phi)$ space (Fig.~\ref{fig:K_fit}a).  Notice that the in-plane anisotropy of NdCo$_5$ is substantially reduced if the $A_6^6\langle r^6 \rangle$ CFP is not taken into account. In fact, without $A_6^6\langle r^6 \rangle$ the single-ion Nd anisotropy is of easy-cone type, in disagreement with the easy-plane observed experimentally.  Hence, the azimuthal magnetic anisotropy of Nd in this compound is very sensitive to the high-rank "66" CFP .  In contrast, the dependence of $E_{anis}$ on the polar angle $\phi$ is rather weak. This implies that the polar dependence of the anisotropy is not a reliable signature of the relative magnitude of $A_6^6\langle r^6 \rangle$.

As shown in Fig.~\ref{fig:K_fit}a, the calculated RE anisotropy energy $ E_{anis}(\theta, \phi)$ can be reasonably well fitted by three anisotropy constants, $K_1$, $K_2$ and $K_3'$, in eq.~\ref{eq:E_anis}. Although a more precise fitting is obtained by including $K_3$, we neglected it to facilitate the comparison with previous experimental measurements, in which $K_3$ has  also been neglected. 
The resulting values of $K_i$ are listed in Table~\ref{table:Ks}. The calculated anisotropy constants are in overall good agreement with experiments, taking into account the large dispersion of experimental values.  In particular, both our theory and experiment find a large negative value of $K_1$ and a  positive constant $K_2$ of smaller magnitude. The overall negative MAE of NdCo$_5$, defined as $E(\vec{M}||a)-E(\vec{M}||c)$, is well  reproduced when the "66" CFP  is taken into account; without this high-rank CFP the magnitude of MAE is severely underestimated. 

The spread  of experimental values is mainly related to uncertainties in extracting $K_i$ values from magnetization data, i.~e., to a two-sub-lattice  model assumed in the analysis. In particular, Ref.~\onlinecite{Ermolenko1980} employed a model allowing for a misalignment of the RE and Co magnetizations with distinct anisotropy constants for each sublattice. In contrast, Refs.~\onlinecite{Klein1975,Ermolenko1976} employed the  Suscksmith-Thompson (ST)\cite{Sucksmith_Thompson} approach to extract the total $K_1$ and $K_2$ values from magnetization curves with the external field applied along the hard direction. This model assumes perfectly aligned Co and RE magnetizations, thus its applicability to two-sublattice systems is questionable\cite{Ermolenko1980}. However, to have a consistent comparison to experimental anisotropy constants  we also extracted them using this approach, by applying an external field $H_c$ along the hard  $\vec{c}$ ([001]) direction.  To that end we minimized the magnetic free energy of NdCo$_5$:
\begin{equation}\label{eq:F_M}
F_M=F_{RE}(\theta_{Co},H_c,T)+K_1^{Co} \sin^2 \theta_{Co} -  \mu_0|\vec{M}_{Co}| H_c \cos \theta_{Co},
\end{equation}
  where second and third terms are the anisotropy and Zeeman energy of the Co sublattice, $\theta_{Co}$ is the azimuthal angle of the Co magnetization $\vec{M}_{Co}$ (confined within the $ac$ plane). The first term is the contribution of Nd sublattice
\begin{equation}\label{eq:F_RE}
F_{RE}(\theta_{Co},H_c,T)=-T \ln \sum_{\Gamma} \exp E_{\Gamma}/T,
\end{equation}
which was calculated from eigenstates $E_{\Gamma}$ of the Hamiltonian (\ref{eq:H_at_full})  with the level positions $\hat{\epsilon}$ (\ref{eq:H_at}) given by the CFPs,  the exchange field $B_{\mathrm{ex}}$ oriented along the direction of Co magnetization, and the external field $H_c$.  We employed our calculated value of 7.5 $\mu_B$ for the total cobalt moment (6.85 $\mu_B$ for the spin moment and 0.65 $\mu_B$ for the orbital moment) and experimental $K_1^{Co}=$ 45 K/(f.u.) measured in YCo$_5$\cite{Alameda1981}. Having found the optimal value of $\theta_{Co}$ we evaluated the azimuthal angle of the total magnetization as a function of $H_c$; then $K_1$ and $K_2$  were computed with the ST formula. The resulting values displayed in parenthesis in Table~\ref{table:Ks} are in a very good agreement with those obtained from experimental data analysis employing the same approach. \cite{TATSUMOTO1971,Ermolenko1976}

These results on the anisotropy constants  can be compared to predictions of the standard linear-in-CF single-multiplet theory for RE magnetic anisotropy in magnetic intermetallics\cite{Kuzmin1995,Kuzmin2007}. In the exchange-dominated regime  $A_6^6\langle r^6 \rangle$ CFP is shown  to contribute only to the polar dependence of  $E_{anis}(\theta, \phi)$,  determined by the anisotropy constant $K_3'$. As follows from (\ref{eq:E_anis}), it should have thus no impact on the average azimuthal ($\theta$) dependence  of  $E_{anis}$, in a drastic disagreement to our numerical results (Fig.~\ref{fig:K_fit}a) showing a strong enhancement of the in-plane anisotropy by the "66" CFP. 

The condition for an exchange-dominated system is given by:
\begin{equation}\label{eq:ex_dominance}
\Delta_{kq}^{CF}=|A_k^q\langle r^q \rangle \Theta_k (\langle \hat{O}_k^q (J)\rangle)_{\mbox{max}}| < J \Delta_{\mathrm{ex}},
\end{equation}
where the exchange splitting $\Delta_{\mathrm{ex}}$ is given by (\ref{H_ex_GSM}), $\Delta_{kq}^{CF}$ is the magnitude of the splitting due to the corresponding $kq$ CF term and the symbol $(\langle \hat{O}_k^q (J)\rangle)_{\mbox{max}}$ designates the largest eigenvalue of the corresponding Stevens operator.  Inserting the calculated values of $A_6^6\langle r^6 \rangle$ and $B_{\mathrm{ex}}$ as well as the appropriate constants for the  GS multiplet $^4I_{9/2}$ of Nd: $J=9/2$, $g_J=8/11$ and $\Theta_6\equiv \gamma_J=-38\cdot 10^{-6}$ and $(\langle \hat{O}_6^6 (J=9/2)\rangle)_{\mbox{max}}=$ 5040 for the Stevens operator $\hat{O}_6^6$ (\ref{eq:Os}) one finds   that the condition of exchange dominance is in fact satisfied for the "66" CFP. The same condition, and even to a larger extent, is satisfied for the "20" CFP. Hence, the failure of the  linear-in-CF theory\cite{Kuzmin1995} can be attributed to its single-multiplet character. The  large "66" CFP  apparently induces strong inter-multiplet effects in NdCo$_5$, as we will demonstrate explicitly in Sec.~\ref{sec:T_dependence} below.

 \begin{figure}[tb]
 	\begin{center}
 		\includegraphics[width=0.90\columnwidth]{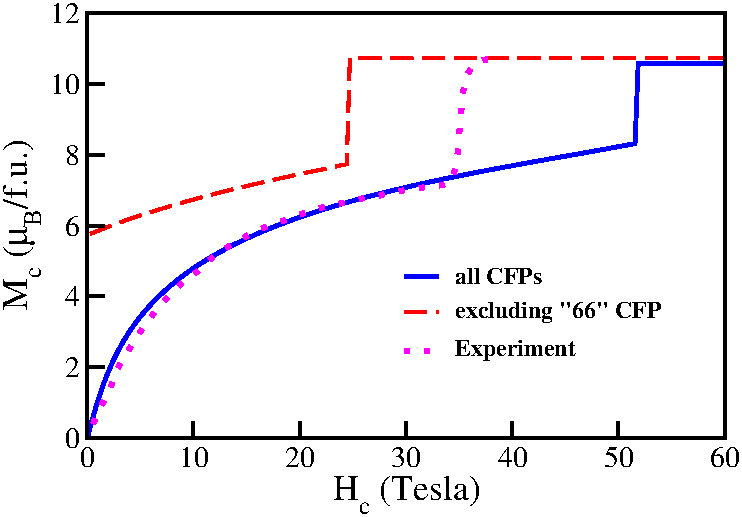}
 	\end{center}
 	\caption{Calculated magnetization along the hard $c$ axis vs. applied field along the same direction at $T=$ 4.2~K. The solid blue and dashed red curves are calculated with and without the "66" CFP, respectively. 
 		Experimental data (dots) are from Ref~\onlinecite{Bartashevich1993}.}
 	\label{fig:Mc_vs_Hc}
 \end{figure}

Using the approach described above, eqs.~\ref{eq:F_M} and \ref{eq:F_RE}, we also calculated the magnetization $M_c$ of NdCo$_5$  along the hard $c$ axis at high external fields $H_c$, up to 60~T, thus simulating the experiments of Refs.~\onlinecite{Bartashevich1993,Zhang1994}.   For the helium temperature we obtain a cube-root-like dependence of  $M_c$ vs. $H_c$ (Fig.~\ref{fig:Mc_vs_Hc}) up to $H^*_c\approx$ 52~T, at which one observe a discontinuous first-order-like jump (i.~e., a first-order magnetization process) to the saturated $M_c$ moment. The theoretical low-field behavior and the saturated total moment of 10.6 $\mu_B$ are in excellent agreement with the experiment (as expected with our  ST anisotropy constants being close to experimental ones). However,  the measured critical field $H^*_c$ is 35~T \cite{Zhang1994,Bartashevich1993}. The overestimation of $H^*_c$ might stem from the approximation of direction independent Co magnetization and Nd-Co exchange coupling used in our calculations  which is questionable\cite{Alameda1981,Ballou1987} and likely to affect our results on the spin-reorientation process at high applied fields.
With the "66" CFP excluded the calculated magnetization curve is qualitatively wrong:  in this case  the easy-cone Nd anisotropy (see Fig.~\ref{fig:K_fit}a) results in a large magnetic moment along the $c$ axis even at zero external field.

\subsection{Temperature dependence of single-ion anisotropy and role of $J$ mixing}\label{sec:T_dependence}

In the previous section we focused on the low-temperature magnetism of NdCo$_5$. Let us now consider  the 4$f$ SIA at elevated temperatures $T$ up to the Curie point ($T_c=$ 910~K) of NdCo$_5$. For a realistic treatment of the RE SIA at high $T$ it is important to take into account the corresponding decrease of $B_{\mathrm{ex}}$ due to a reduced magnetization of the Co sublattice. We thus scaled the zero-temperature value of $B_{\mathrm{ex}}$ with temperature as  $B_{\mathrm{ex}}(T)=B_{\mathrm{ex}}m(\tau)$, where $m(\tau)$ is the reduced Co magnetization $M(T)/M(0)$ as a function of reduced temperature $T/T_c$. For  $m(\tau)$ we employed a semi-empirical formula of Kuz'min~\cite{Kuzmin2005} parametrized for YCo$_5$. Using this $B_{\mathrm{ex}}(T)$  we obtained $E_{anis}(T)=F_{RE}(\theta_{Co}=\pi/2,H_c=0,T)-F_{RE}(\theta_{Co}=\pi/2,H_c=0,T)$ with $F_{RE}$ calculated in accordance with eq.~\ref{eq:F_RE} as detailed above.

\begin{figure}[t!]
	\begin{center}
		\includegraphics[width=0.9\columnwidth]{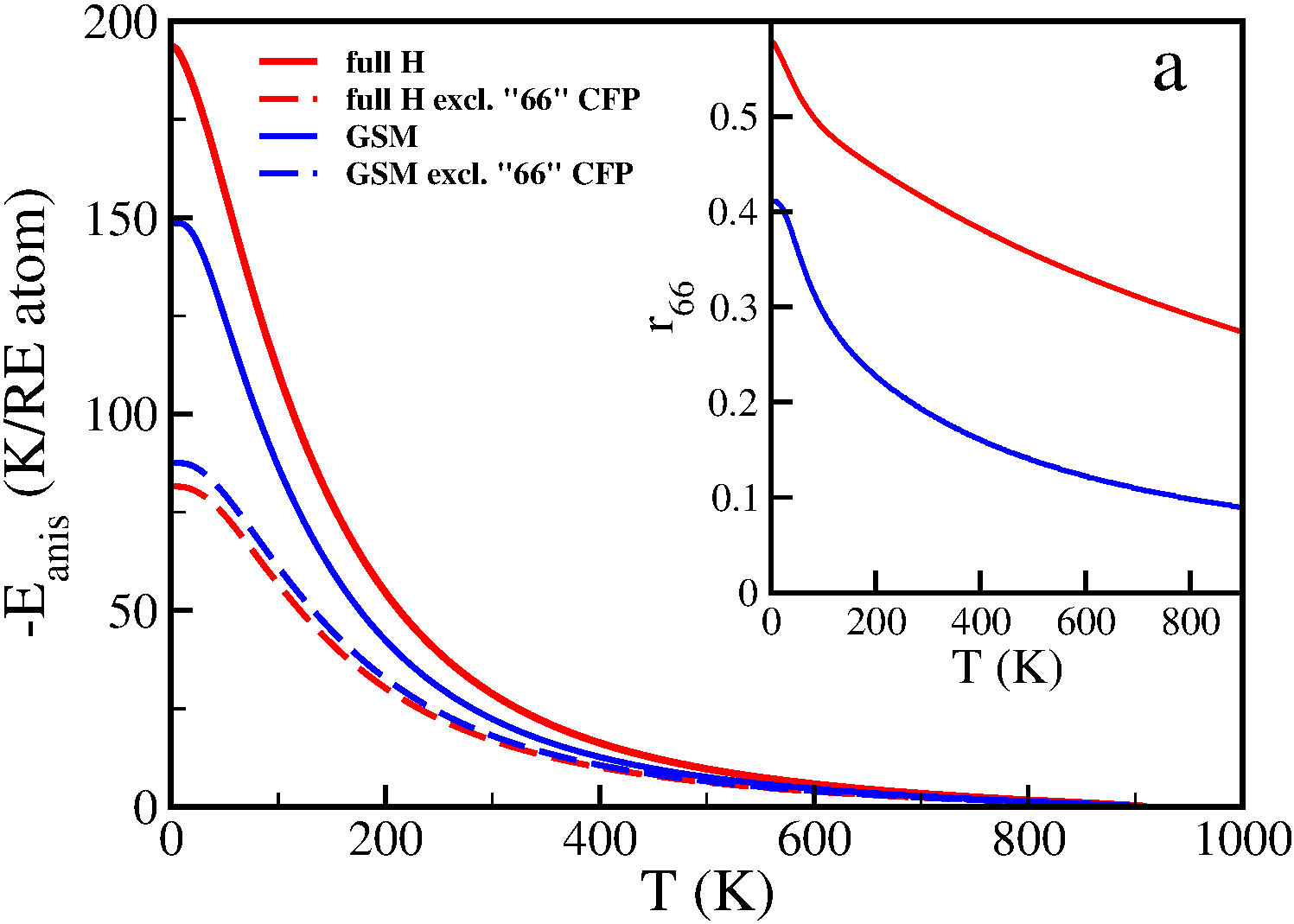}
		\includegraphics[width=0.9\columnwidth]{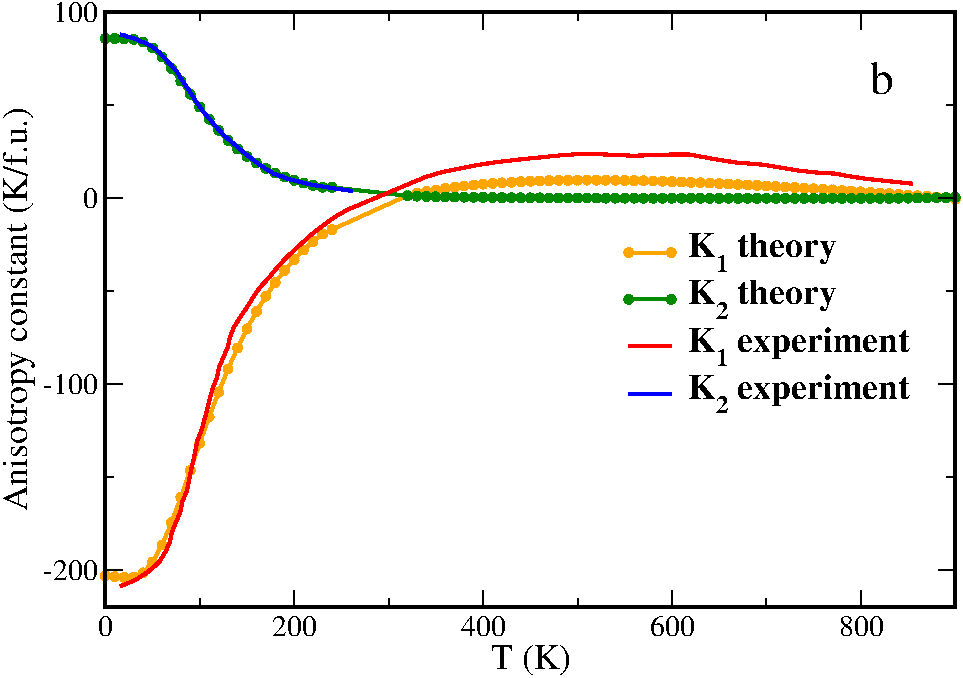}
	\end{center}
	\caption{ a. RE contribution to magnetic anisotropy $E_{anis}$ in NdCo$_5$ vs. temperature. The solid and dashed lines are calculated including all CFPs and with the "66" CFP excluded, respectively. Inset: the relative contribution of the "66" CFP to $E_{anis}$ vs. $T$. b.  Temperature dependence of the  anisotropy constants $K_1$ and $K_2$, evaluated with the ST-method\cite{Sucksmith_Thompson}.  The ST fitting becomes poorly defined close to the spin-reorientation transition of NdCo$_5$; therefore, we de not show the points in its vicinity. The experimental curves are from Ermolenko~\cite{Ermolenko1976}.  }
	\label{fig:E_anis_vs_T}
\end{figure}

The calculated RE anisotropy energy  is plotted in Fig.~\ref{fig:E_anis_vs_T}a. As expected $E_{anis}$ exhibits a rapid decrease with increasing temperature. More interestingly, by comparing $E_{anis}$ calculated with and without the "66" CFP one concludes that its strong impact on the anisotropy persists in the high-temperature regime. Indeed, its relative contribution $r_{66}=(E_{anis}-\tilde{E}_{anis})/E_{anis}$, where $\tilde{E}_{anis}$ is calculated excluding the "66" CFP, decreases rather slowly with temperature and is still about 27\% near $T_c$ (red curve in the inset of Fig.~\ref{fig:E_anis_vs_T}a).

This behavior is quite unexpected. In fact, the high-temperature expansion of the RE single-ion anisotropy (see, e.g., Ref~\onlinecite{Kuzmin1995,Kuzmin2007}) predicts that only the "20" CFP  contributes to the MAE in the leading order in $1/T$.  Within this single-multiplet formalism 
higher-rank 
CFPs are found to contribute only to higher orders in $1/T$ and should become relatively unimportant at high $T$ approaching $T_c$. This conclusion follows from orthogonality properties of the Stevens and angular moment operators and should hold even at relatively large values of high-rank CF contributions, as far as they are smaller than $T$.

In order to better  understand the origin of this behavior we computed the temperature evolution of $E_{anis}$ and $\tilde{E}_{anis}$  
using the Stevens formalism, eqs.~\ref{eq:H_cf_St} and \ref{H_ex_GSM}, i.e. including only the GSM. 
One sees that excluding excited multiplets reduces the contribution of "66" CFP by about a quarter at $T=$ 0 and by about 60\% at  $T=$ 300~K (cf. the red and blue curves in inset of Fig. 7a, which give the contribution of "66" with and without the excited multiplets, respectively). 
The inter-multiplet mixing thus significantly increases the "66" CFP  contribution to the anisotropy, particularly, at room temperature and above.  Inversely, the role of inter-multiplet mixing is drastically enhanced by this CFP. Indeed, with the "66" CFP excluded the single-multiplet and full calculations produce very similar values for the RE anisotropy energy (Fig.~\ref{fig:E_anis_vs_T}).

We have also evaluated the temperature dependence of the anisotropy constants $K_1$ and $K_2$ using the ST approach, as was employed by Ermolenko~\cite{Ermolenko1976} to extract the anisotropy constants from experimental magnetization curves. The agreement of our theoretical $K_i(T)$, calculated with all CFPs included, with experimental data is excellent, in particular at low and intermediate temperatures.

\subsection{Comparison to TbCo$_5$}

Let us now turn to the case of heavy-RE "1-5" system TbCo$_5$. The CFPs of Tb obtained by the self-interaction suppressed DFT+HubI method (Fig.~\ref{fig:CF} and Appendix Table~\ref{table:CFP_comparison}) are qualitatively similar to those of Nd presented above. The negative value  -118~K of low-rank CFP $A_2^0\langle r^2 \rangle$ indicates in-plane Tb SIA in this compound, similarly to NdCo$_5$, but its magnitude is noticeably smaller. The magnitude of "66" CFP is quite large, 440~K, but is almost 3 times smaller than in NdCo$_5$. The ratio of these two CFPs, $\frac{A_6^6\langle r^6 \rangle}{A_2^0\langle r^2 \rangle}$, is almost the same in TbCo$_5$ and NdCo$_5$, seemingly suggesting an equally strong impact of the  "66" CFP in these systems.

We performed the same calculation of the anisotropy energy as a function of  $\theta$ and $\phi$ for  TbCo$_5$ as for NdCo$_5$ and then extracted the values of anisotropy constants $K_1$, $K_2$ and $K_3'$. As shown in  Fig.~\ref{fig:K_fit}, with the "66" CFP included, the easy direction lies along  the hexagonal $a$ axis( $\theta=\pi/2$, $\phi=0$). The absolute value of the single-ion contribution to MAE,  $E_{RE}(\vec{M}||a)-E_{RE}(\vec{M}||c)=-$106~K, is about 2 times smaller in TbCo$_5$ than that of NdCo$_5$. 

The calculated anisotropy constants are listed in Table~\ref{table:Ks}. In contrast to NdCo$_5$ we obtain negative values for Tb $K_1$ and $K_2$, which are of comparable magnitude. The overall MAE (including the Co contribution)  is negative, corresponding to in-plane $a$ easy axis, and it is about twice smaller than in NdCo$_5$. These findings are in qualitative agreement with the measurements of Ermolenko~\cite{Ermolenko1980}, which is the only experimental work, to our awareness, reporting the low-temperature anisotropy constants of TbCo$_5$. Our calculated $K_1$ anisotropy constant and, correspondingly, MAE seem to be  underestimated, if compared to Ref.~\onlinecite{Ermolenko1980}. However, as already mentioned above, this work employed a non-standard approach for extracting anisotropy constants. The RE anisotropy constant $K_1$ of NdCo$_5$ reported by Ermolenko is also overestimated compared to other experimental references. 
  
Our calculated GS wave function of Tb 4$f^8$ shell, defined in the same coordinate frame as the Nd GS wave function \ref{eq:Nd_GS}, is  the pure total moment eigenstate:
  \begin{equation}\label{eq:Tb_GS}
  \Psi_{GS}^{Tb}=|66\rangle,
  \end{equation}
  corresponding to the fully saturated Tb moment (see Table ~\ref{table:WFs_TbCo5} in the Appendix for a full level scheme). 
   \begin{figure}[ht]
   	\begin{center}
   		\includegraphics[width=0.80\columnwidth]{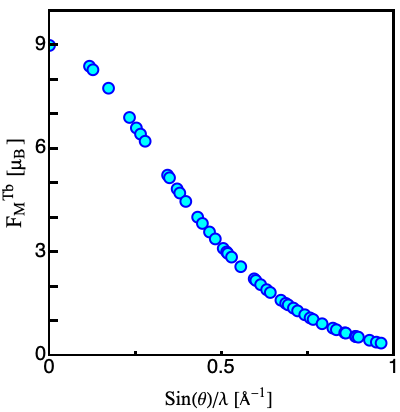}
   	\end{center}
   	\caption{Tb magnetic structure factors in TbCo$_5$ predicted from converged GS   $\Psi_{GS}^{Tb}$ computed at the same reciprocal lattice vectors $\vec{\varkappa}$ as those in the PNS experiment on NdCo$_5$ \cite{Alameda1982}.} 
   	\label{fig:MDSF_TbCo5}
   \end{figure}
%
Only a negligible change in the GS is observed with the "66" CFP excluded, which becomes $0.999|6+6\rangle+0.045|6+4\rangle$, the splitting to the first excited state (almost pure $|65\rangle$ in the both cases) then decreases from 232 to 217~K. %
Fig.~\ref{fig:MDSF_TbCo5} shows the Tb contribution to the neutron magnetic structure factor, $F_M$, of TbCo$_5$ predicted from this GS. It shows no anisotropy.
  
Therefore, we conclude that  $A_6^6\langle r^6 \rangle$ does not affect the low-temperature magnetism of Tb and  has a rather insignificant  impact on its magnetic anisotropy, other then inducing, obviously, some planar anisotropy. This behavior is in sharp contrast to that of NdCo$_5$, what might seem to be in contradiction to approximately the same relative value $A_6^6\langle r^6 \rangle$, with respect to $A_2^0\langle r^2\rangle$,  in these two systems. However, the Stevens factor $\gamma_J = -1.121\cdot 10^{-6}$ for the  GS multiplet $^7F_6$ of Tb is much smaller than that for  Nd $^4I_{9/2}$. The relative importance  of  "20" and "66" terms in (\ref{eq:H_cf_St}) may be estimated from the ratio of splittings (\ref{eq:ex_dominance}) generated by each CFP in a given GS multiplet: 
  \begin{equation}
  \label{eq:d_ratio}
  d=\frac{\Delta_{20}^{CF}}{\Delta_{66}^{CF}}=\frac{\gamma_J A_6^6\langle r^6 \rangle(\langle \hat{O}_6^6(J)\rangle)_{\mbox{max}}}{\alpha_J A_2^0\langle r^2 \rangle \rangle (\langle \hat{O}_2^0(J)\rangle)_{\mbox{max}}}.
  \end{equation}
       Evaluating (\ref{eq:d_ratio})  with our calculated CFPs we find $d=$ 3.28 and 0.19 for  Nd and Tb, respectively, the "66" CFP being thus about 17 times more significant in the former case. Therefore, while our calculations predict a large "66" CFP in all RECo$_5$ compounds calculated so far, the impact of this CFP on RE magnetic moment and anisotropy is ion-dependent. This impact is expected to be particularly significant in light RE ions, for which the rank-6 Stevens factor  $\gamma_J$ is relatively large and rather weak in heavy RE with large GS $J$, like Tb or Dy. 
       
       Moreover, the Tb CF states within its GS multiplet feature much smaller $J$-mixing as compared to the Nd ones (see Tables~\ref{table:WFs_NdCo5} and \ref{table:WFs_TbCo5} in Appendix). Hence, in contrast to the Nd case, no strong impact of $J$-mixing on the anisotropy is expected.
    
\section{Analysis: electronic structure, hybridization and rank-6 crystal-field in RECo$_5$}\label{sec:analysis}

 \begin{figure*}
 	\begin{center}
 		\includegraphics[width=2.0\columnwidth]{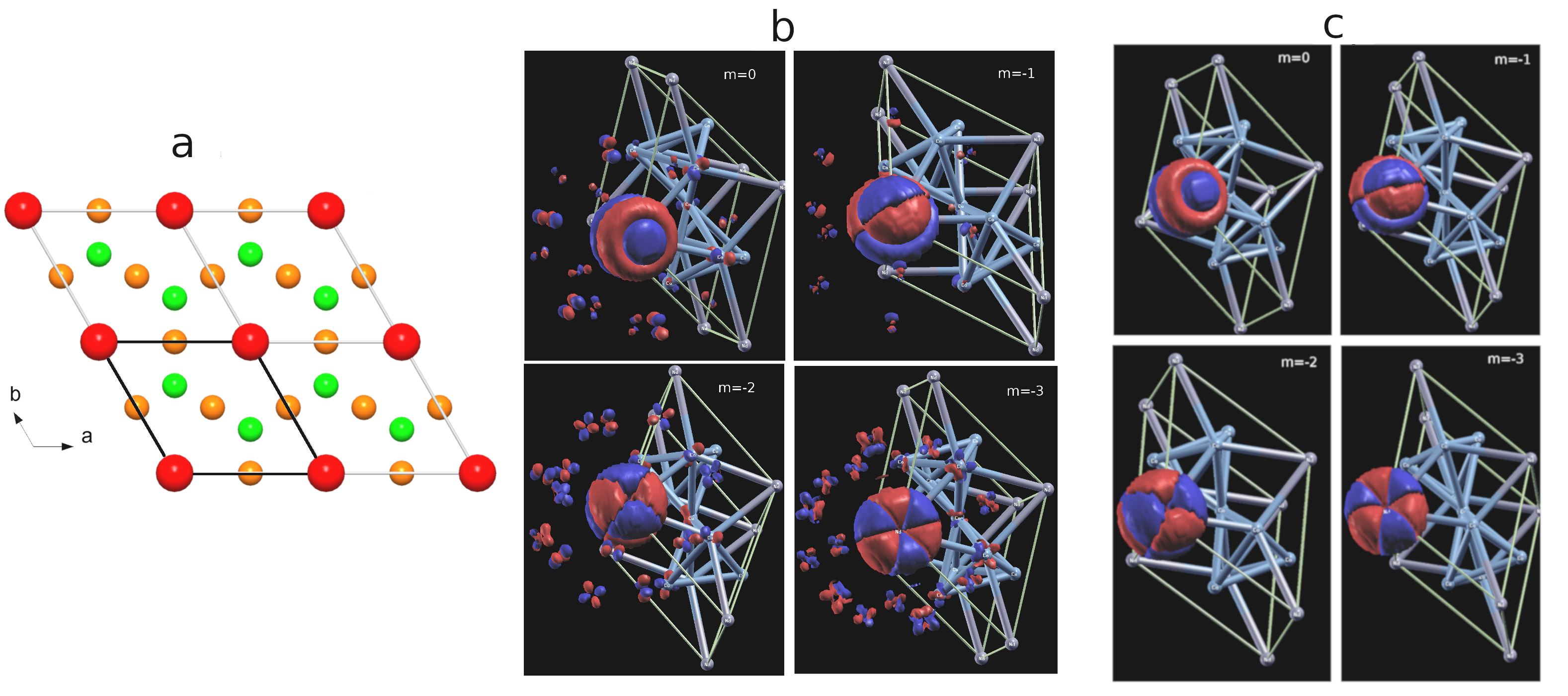}
 	\end{center}
 	\caption{a. RECo$_5$ crystal structure (view along the hexagonal [001] direction).  Red, green and orange balls depict RE, Co 2$c$ and Co 3$g$ sites, respectively; the unit cell is indicated by bold black lines.  b.   Nd 4$f$ Wannier orbitals for $m=$0, -1 (upper row), -2, -3 (bottom row) constructed using the small energy window  $\mathcal{W}_s \in$[-2:2]~eV . c The same orbitals constructed using  the large energy window  $\mathcal{W}_l \in$[-10:10]~eV.}
 	\label{fig:WF}
 \end{figure*}

 As shown in Fig.~\ref{fig:CF} above, the present DFT+HubI method  predicts an unexpectedly large value of $A_6^6\langle r^6 \rangle$  in all three  RECo$_5$ compounds studied to date ($RE=$Nd, Sm, Tb).  In addition, the magnitude of this CFP seems to reduce along the series, being the largest in  Nd and smallest in Tb. In this section we aim at identifying physical origins of these results. 
 
 The crystalline environment of RE site in RECo$_5$ is  invariant under a 6-fold rotation (Fig.~\ref{fig:WF}a), but not under an arbitrary rotation about the $c$ axis. This is precisely the symmetry of $A_6^6\langle r^6 \rangle\hat{O}_6^6$ term, which is invariant under the 6-fold  rotation about the $c$ hexagonal axis. This points out to its likely origin in a spatially non-uniform in-plane interaction between $R$ and its Co neighbors. The main contribution to the "66" CFP  is apparently missed by open-core approaches (see Table~\ref{table:CFPSs_comp}). This suggests hybridization  between RE and  Co states as a likely origin of the large "66" CFP.  The symmetry of hybridization is determined by the local environment of RE ions. Mixing of localized  4$f$s with, for example,  Co 3$d$  states, 
  which are also to some degree localized, 
  should lead in a simple tight-binding picture to the formation of directed bonds leading to the expected 6-fold symmetry of the resulting CF contribution.

 \begin{figure*}[t!]
 	\begin{center}
 		\includegraphics[width=2.0\columnwidth]{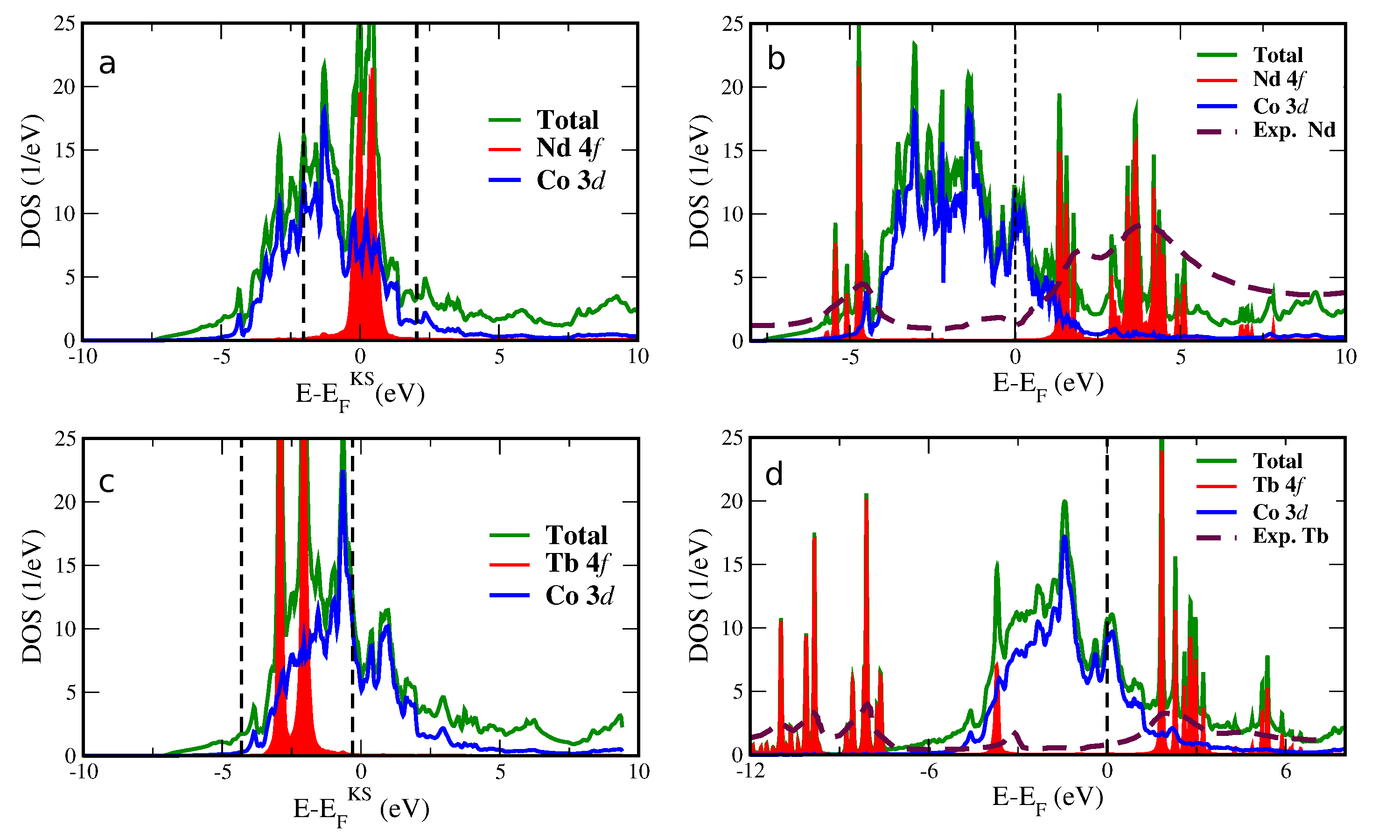}
 	\end{center}
 	\caption{a. Density of Kohn-Sham (KS) states in NdCo$_5$ as obtained from the charge density converged in DFT+HubI. The large window $\mathcal{W}_l \in$[-10:10]~eV includes all states shown on this plot. 
 		The range included into  the small energy window  $\mathcal{W}_s \in$[-2:2]~eV is indicated by vertical dashed lines. b. The DFT+HubI spectral function of NdCo$_5$ (calculated with the small window $\mathcal{W}_s$). The Nd 4$f$ spectral function features sharp peaks corresponding to transitions between atomic multiplets. The same plots for TbCo$_5$ are shown in panels c and d, respectively. Notice the shift of the Tb 4$f$ KS band to lower energies. The experimental photoemission and inverse-photoemission spectra  displayed as brown dashed line in b and d are for the Nd and Tb metals\cite{Lang1981}.
 	}
 	\label{fig:DOS}
 \end{figure*}
 
 These qualitative arguments can be verified within the present DFT+HubI approach by exploiting the flexibility of its 4$f$-orbitals basis.  As hybridization effects are not included explicitly into the local 4$f$ problem  within the Hubbard-I approximation, they can only  implicitly enter into (\ref{eq:H_at_full}), through the shape of  4$f$ orbitals in which matrix elements $\langle \hat{H}_{KS} \rangle^{ff}$ in (\ref{H1el_HI}) are evaluated.  4$f$ orbitals in the present framework are Wannier orbitals (WO) constructed using the projective two-step approach of Refs.~\onlinecite{Amadon2008,Aichhorn2009}. First, an  initial 4$f$ basis is generated by expanding the 4$f$ local orbitals $|\chi_{m\sigma}\rangle$ defined within RE "atomic sphere" in terms of the Bloch states $|\psi_{\vk\nu}\rangle$  enclosed within a chosen energy window $\mathcal{W}$:
 $$
 |\chi^{\vk}_{m\sigma}\rangle=\sum_{\nu \in \mathcal{W}} |\psi_{\vk\nu}\rangle \langle \psi_{\vk\nu} | \chi_{m\sigma}\rangle.
 $$
 The resulting set of orbitals $ |\chi^{\vk}_{m\sigma}\rangle$ is not orthonormal due to the  incompleteness of the Bloch basis restricted by the range $\mathcal{W}$. Subsequent orthonormalization of this initial set  leads to a true Wannier basis $\{\mathcal{\omega}_{m\sigma}\}$, with the resulting orbitals extending beyond RE site due to hybridization mixing of 4$f$ states with other bands. Using a large  $\mathcal{W}_l$ range reduces the degree of incompleteness of the Bloch basis; the set $\{\mathcal{\omega}_{m\sigma}\}$ in this case does not  differ much from initial  $\{\chi_{m\sigma}\}$. With this basis choice  DFT+HubI calculations  are expected to produce results similar  to those of the open-core framework. The narrow $\mathcal{W}_s$ range, enclosing mainly 4$f$ bands, results in extended WO due to hybridization admixture of other characters to those bands, as shown by Delange~{\it et al.}~\cite{Delange2017} on the example of "1-12" intermetallics. The matrix elements $\langle \hat{H}_{KS} \rangle^{ff}$ computed in such an extended WO basis are affected by hybridization. 
 
  \begin{table}[b]
  	\caption{\label{table:CFPlw} Calculated crystal-field parameters (in K) and exchange field (in Tesla) in NdCo$_5$ using the large $\mathcal{W}_l$ and small $\mathcal{W}_s$ energy windows as well as intermediate windows $[-2:10]$ and $[-10:2]$. }
  	\centering
  	 	\begin{ruledtabular}
  		\begin{tabular}{l  l  l  l  l  l}
  		Energy window (eV) & $A_2^0\langle r^2 \rangle$ &$A_4^0\langle r^4 \rangle$ & $A_6^0\langle r^6 \rangle$ &  $A_6^6\langle r^6 \rangle$ & $B_{\mathrm{ex}}$ \\
  		\hline
  		$[-10:10]$, ($\mathcal{W}_l$)& -198 & -57 & 1 & 45 & 326 \\
  		$[-2:10]$ & -388 & -50 & 7 & 357 & 332  \\
        $[-10:2]$ & -125 & -34 & 19 & 731 & 287  \\
  		$[-2:2]$,  ($\mathcal{W}_s$) & -285 & -33 & 36 & 1134 & 292  \\
  	\end{tabular}
\end{ruledtabular}
  \end{table}
 
 We have performed test calculations for NdCo$_5$ employing the large window $\mathcal{W}_l \in$[-10:10]~eV, containing all Co 3$d$ and a large part of Nd 5$d$ states (see\footnote{In our DFT+HubI calculations the  exchange field on the RE 4$f$ shell (i.~e. $B_{\mathrm{ex}}$) is due to the Co spin polarization only, as the 4$f$ own magnetization density is suppressed by averaging, see Sec.~\ref{subsec:elstruct}. The value of $B_{\mathrm{ex}}$ (Table~\ref{table:CFP_comparison}) is small compared to the width of RE KS 4$f$ band, which remains, correspondingly, essentially not spin-polarized, as is seen in Figs.~\ref{fig:DOS}a and c.} Fig.~\ref{fig:DOS}a). As noted in Sec.~\ref{sec:method}, the rest of NdCo$_5$ calculations in this work employed  extended WO constructed using the window $\mathcal{W}_s \in$[-2:2]~eV around the Kohn-Sham Fermi energy $E_F^{KS}$. As one sees in Fig.~\ref{fig:DOS}a, $\mathcal{W}_s$ includes all Nd 4$f$, whereas part of Co 3$d$ and almost all Nd 5$d$ are excluded. The impact of hybridization on the resulting WO can be qualitatively analyzed by plotting them in the real space. The Nd 4$f$ orbitals  in NdCo$_5$  constructed for different magnetic quantum number $m$ by using the large and small energy windows are depicted in Fig.~\ref{fig:WF}b and \ref{fig:WF}c, respectively. The WO on this plot were constructed neglecting the spin-orbit coupling in order to highlight the orbital dependence of their spread.   The same value is used to define the isodensity surface in the both cases. 
 
 One sees that the small-window WO are extended and leak to neighboring Co sites. This leakage is orbital-dependent (being rather small for $m=$ -1 and large for $m=$ -2 and -3), hence, it directly contributes to the splitting of the corresponding one-electron levels.  In contrast, the "large-window" WO exhibit no leakage to the Co neighbors (see Fig.~\ref{fig:WF}). Therefore, the CFPs calculated in this case using DFT+HubI approach do not include any contribution of hybridization and will be determined solely by the electrostatic contribution.

We carried out full DFT+HubI crystal-field calculations using the large energy window $\mathcal{W}_l \in$[-10:10]~eV for constructing localized WO; all other parameters of these calculations are identical to those using with the small window $\mathcal{W}_s$. The CFPs and $B_{\mathrm{ex}}$ obtained with the two choices for WO are compared in Table~\ref{table:CFPlw}. One observes a very small impact on $B_{\mathrm{ex}}$ and some decrease in the magnitude of the low-rank "20" CFP. In contrast, the value of  $A_6^6\langle r^6 \rangle$ is reduced by a factor of 25  when the localized WO (constructed using $\mathcal{W}_l$) are employed.  
Not surprisingly, with such a small "66" CFP a fully-polarized Nd GS of almost pure $|9/2;-9/2\rangle$ is obtained. From this analysis we conclude that the crucial large "66" CFP in NdCo$_5$ and in RECo$_5$ in general, is due to hybridization effects, the purely electrostatic contribution being is quite insignificant. 

We have also performed calculations with the window extended either to include only occupied valence states, $[-10:2]$~eV, or  a wide range of unoccupied states, $[-2:10]$~eV. As compared to the localized WO ($\mathcal{W}_l$), these WO effectively include the hybridization with empty and filled states, respectively. The resulting zonal "20" CFPs (Table~\ref{table:CFPlw}) exhibits a non-monotonous dependence on the window size, apparently indicating hybridization contributions of different signs stemming from filled and empty states. In contrast, the "66" CFP strongly increases in both cases, but the impact of hybridization with empty states (RE 5$d$, Co 4$s$) is noticeably more pronounced. 

The KS electronic structure of TbCo$_5$,  obtained from converged DFT+HubI calculations, is displayed in Fig.~\ref{fig:DOS}c. Tb 4$f$ bands are located significantly lower in energy as compared to Nd 4$f$ bands in NdCo$_5$. Such evolution along the RE series is generally expected.  Therefore, as Tb 4$f$ KS bands are not anymore pinned at $E_F^{KS}$, we  continuously adjusted the position of $\mathcal{W}_s$ in the course of DFT+HubI calculation,  see the Method section. 

In Figs.~\ref{fig:DOS}b and d we display the calculated DFT+HubI spectral function for NdCo$_5$ and TbCo$_5$, respectively. The quasi-atomic multiplet structure of RE 4$f$ is compared to experimental photoemission spectra (PES) and inverse PES  of the Nd and Tb metals\cite{Lang1981} (we are not aware of  any PES experiments on Nd and Tb "1-5" systems). One observes a very good agreement between the positions of 4$f$ peaks in DFT+Hub-I and experimental PES. Notice that, in contrast to the previous DFT+HubI calculations of Refs.~\onlinecite{Granas2012,Locht2016}, we did not adjust the position of the occupied RE 4$f$ states  to that in experimental PES. Although the multiplet structure and  the splitting between empty and  occupied 4$f$ states are mainly determined by the input local Coulomb interaction, the position of the 4$f$ states center-weight  relative to other bands is determined by that of the  KS 4$f$ bands (Figs.~\ref{fig:DOS}a and c). The latter comes out of our charge self-consistent DFT+Hub-I calculations, which, therefore, predict quantitatively correctly the lower position of the Tb 4$f$ band as compared to the Nd one (cf. the position of 4$f$ band relative to KS $E_F$ in NdCo$_5$ and TbCo$_5$, Figs.~\ref{fig:DOS}a and c, respectively).

As described above for the case of NdCo$_5$, the principal contribution to the "66" CFP is due to the hybridization between RE 4$f$ and empty conduction bands. The predicted shift of the Tb KS 4$f$ states to lower energy  should weaken this hybridization, hence the observed reduction of the "66" CFP in TbCo$_5$ as compared to the case of Nd. On the basis of this argument one expects  a decrease of "66" CFP  in RECo$_5$ along the RE series, which we indeed find, see Fig.~\ref{fig:CF}.

\section{Conclusions}

We have calculated crystal-field parameters (CFPs) and rare-earth single-ion magnetic anisotropy in ferrimagnetic intermetallics NdCo$_5$ and TbCo$_5$  using the {\it ab initio} DFT+Hubbard-I methodology of Ref.~\onlinecite{Delange2017}.  Our study reveals that the order-six CFP "66" $A_6^6\langle r^6 \rangle$ takes exceptionally large values in these RECo$_5$ systems (as well as in SmCo$_5$ calculated before in Ref.~\onlinecite{Delange2017}), especially in the light RE element Nd. In the present work we aimed at evaluating the impact of this large order-six CFP on RE magnetization and single-ion anisotropy.  In particular, in NdCo$_5$,  this CFP is found to freeze the ground-state Nd moment well below its fully saturated value. We show that this freezing of the GS moment, previously observed\cite{Alameda1982} but not explained, represents in fact an experimental fingerprint of a large $A_6^6\langle r^6 \rangle$ CFP in this system. Our calculations reveal a strong  impact of this CFP on the NdCo$_5$ anisotropy and its temperature dependence; the calculated anisotropy constants are in quantitative agreement with experimental data. Our calculations also predict a large value of this CFP in TbCo$_5$, which is, however, not as huge as that  of NdCo$_5$. Moreover, in the case of TbCo$_5$ the "66" CFP has a very weak influence on the magnetic anisotropy and does not affect the GS magnetization. This is explained by a relatively small order-six Stevens coefficient of the Tb GSM reducing the impact of order-six CFPs on its magnetism.  The influence of $A_6^6\langle r^6 \rangle$ on the magnetism of RECo$_5$ is thus RE-ion-specific.

The large value of $A_6^6\langle r^6 \rangle$ in RECo$_5$ is shown to be induced by hybridization between the RE 4$f$ shell and its 6-fold coordinated crystalline environment. In our DFT+Hubbard-I  approach this hybridization is taken into account indirectly, through the shape of 4$f$ orbitals, which become less localized due to hybridization effects. Using the flexibility of our orbital basis we clearly demonstrate that by neglecting the impact of hybridization to CFPs one reduces the magnitude of calculated $A_6^6\langle r^6 \rangle$ by more than one order. The hybridization with empty itinerant states is shown to be the most important contribution into the "66" CFP. The progressive shift of 4$f$ states to lower energies along the RE series reduces this hybridization resulting in a progressive reduction of the "66" CFP from NdCo$_5$ to TbCo$_5$.  

More generally, this work shows that hybridization mixing of RE 4$f$ shell with its $q$-fold coordinated environment may lead to the appearance of large CFPs $A_k^q\langle r^k \rangle$, with $q\ne 0$. These high-order CFPs are traditionally considered to be much less important for the RE single-ion magnetic anisotropy as compared to low-order $A_2^0\langle r^2 \rangle$. The present work shows that this assumption does not always hold. The local environment of a RE ion can be modified with TM substitutions or small-atom insertions changing the hybridization of RE 4$f$ with other bands, and, hence,  these high-order CFPs. As shown in the present work, by using an advanced  {\it ab initio} methodology one can quantitatively describe  such hybridization-induced CFPs and their impact on the magnetocrystalline anisotropy. This opens an opportunity for theoretical optimization of RE-TM intermetallics with respect to such properties as the single-ion magnetic anisotropy, the spin-reorientation transition temperature, or the magnetocaloric effect. 

\section*{Acknowledgments}

L.~P. acknowledges useful discussions with S.~Khmelevski and T.~Miyake. This work was supported by the European Research Council grants ERC-319286-"QMAC" and ERC-617196-"CorrelMat", as well as the DFG-ANR grant "RE-MAP". We also acknowledge the support by the future pioneering program
"Development of magnetic material technology for high-efficiency motors" (MagHEM),
grant number JPNP14015, commissioned
by the New Energy and Industrial Technology Development Organization (NEDO). We are  grateful to the computer team at CPHT for support.

\newpage
\appendix 
 
\renewcommand{\arraystretch}{1.15}
\begin{table*}[h]
\onecolumngrid
\centering
\caption{\label{table:WFs_NdCo5} Calculated eigenvalues and eigenstates of Nd 4$f$ shell in NdCo$_5$}
\begin{tabular}{ p{2.5cm}   l }
	\hline\hline
	$E-E_{GS}$, K {     }& Eigenstates in $|JM\rangle$ basis \\ \hline
	0 & $+0.827|9/2-9/2\rangle-0.536|9/2-5/2\rangle-0.096|11/2-9/2\rangle+0.094|11/2-5/2\rangle-0.089|9/2-1/2\rangle$     \vspace{1mm} \\
	220 & $+0.702|9/2-3/2\rangle+0.690|9/2-7/2\rangle-0.117|9/2+5/2\rangle-0.103|11/2-3/2\rangle-0.063|9/2+1/2\rangle$  \vspace{1mm} \\
	280 & $+0.760|9/2-5/2\rangle+0.535|9/2-9/2\rangle+0.305|9/2-1/2\rangle-0.158|9/2+3/2\rangle-0.092|9/2+7/2\rangle$ \\
	& $-0.079|11/2-1/2\rangle-0.045|11/2-5/2\rangle+0.032|11/2+7/2\rangle$  \vspace{1mm} \\
	526 & $+0.708|9/2-7/2\rangle-0.687|9/2-3/2\rangle+0.091|11/2-3/2\rangle+0.081|9/2+1/2\rangle-0.079|9/2+5/2\rangle$ \\
	& $+0.058|9/2+9/2\rangle-0.034|11/2-7/2\rangle-0.032|11/2-11/2\rangle$  \vspace{1mm} \\
	642 & $+0.668|9/2-1/2\rangle-0.613|9/2+3/2\rangle-0.333|9/2-5/2\rangle-0.189|9/2+7/2\rangle-0.138|9/2-9/2\rangle$ \\
	& $-0.087|11/2-1/2\rangle+0.056|11/2+3/2\rangle+0.036|11/2+7/2\rangle$  \vspace{1mm} \\
	697 & $+0.789|9/2+5/2\rangle+0.567|9/2+1/2\rangle-0.183|9/2+9/2\rangle+0.107|9/2-7/2\rangle+0.068|9/2-3/2\rangle$ \\
	& $-0.046|11/2+5/2\rangle-0.040|11/2+9/2\rangle-0.032|11/2-3/2\rangle$  \vspace{1mm} \\
	738 & $+0.666|9/2+3/2\rangle+0.653|9/2-1/2\rangle+0.330|9/2+7/2\rangle-0.094|9/2-5/2\rangle-0.078|11/2-1/2\rangle$ \\
	& $-0.056|11/2+7/2\rangle$  \vspace{1mm} \\
	829 & $+0.807|9/2+1/2\rangle-0.524|9/2+5/2\rangle+0.201|9/2+9/2\rangle-0.102|11/2+1/2\rangle-0.094|9/2-7/2\rangle$ \\
	& $+0.075|11/2+5/2\rangle+0.071|9/2-3/2\rangle-0.040|11/2-3/2\rangle$  \vspace{1mm} \\
	1070 & $+0.956|9/2+9/2\rangle+0.252|9/2+5/2\rangle-0.102|11/2+5/2\rangle-0.064|9/2+1/2\rangle+0.055|9/2-3/2\rangle$ \\
	& $-0.055|11/2+9/2\rangle$ \vspace{1mm} \\ 
	1111 & $+0.905|9/2+7/2\rangle-0.387|9/2+3/2\rangle-0.139|11/2+7/2\rangle-0.065|9/2-1/2\rangle+0.059|9/2-5/2\rangle$ \\
	& $+0.041|11/2+11/2\rangle+0.040|11/2+3/2\rangle$ \vspace{1mm} \\
	\hline		
\end{tabular}
\end{table*}

\begin{table*}[h]
	\onecolumngrid
	\centering
	\caption{\label{table:WFs_TbCo5} Calculated eigenvalues and eigenstates of Tb 4$f$ shell in TbCo$_5$}
	
	\begin{tabular}{ p{2.5cm}   l }
		\hline\hline
		$E-E_{GS}$, K & Eigenstates in $|JM\rangle$ basis \\ \hline
		
		 0 & $+1.000|6+6\rangle$  \\

    232 & $+0.994|6+5\rangle+0.091|6+3\rangle+0.048|5+5\rangle$  \\

    428 & $+0.991|6+4\rangle+0.098|6+2\rangle+0.080|5+4\rangle$  \\

    634 & $+0.989|6+3\rangle-0.093|6+5\rangle+0.088|5+3\rangle+0.076|6+1\rangle$  \\

    844 & $+0.988|6+2\rangle-0.098|6+4\rangle+0.095|5+2\rangle+0.059|6+0\rangle$  \\

   1050 & $+0.989|6+1\rangle+0.103|5+1\rangle-0.076|6+3\rangle+0.060|6-1\rangle$  \\

   1251 & $+0.989|6+0\rangle+0.109|5+0\rangle+0.076|6-2\rangle-0.060|6+2\rangle$  \\

   1448 & $+0.987|6-1\rangle+0.110|5-1\rangle+0.090|6-3\rangle-0.062|6+1\rangle$  \\

   1647 & $+0.987|6-2\rangle+0.104|5-2\rangle+0.088|6-4\rangle-0.078|6+0\rangle$  \\

   1852 & $+0.989|6-3\rangle+0.093|5-3\rangle-0.091|6-1\rangle+0.063|6-5\rangle$  \\

   2059 & $+0.992|6-4\rangle-0.087|6-2\rangle+0.083|5-4\rangle$  \\

   2260 & $+0.995|6-5\rangle+0.071|5-5\rangle-0.061|6-3\rangle$  \\

   2440 & $+1.000|6-6\rangle$  \\

		\hline
		
	\end{tabular}  
\end{table*}

\begin{table}[t!]
	\caption{\label{table:CFP_comparison} Calculated crystal-field parameters (in K) and exchange field (in Tesla) in RECo$_5$ ($R$=Nd,Tb). The quantization axis is along the hexagonal [001] direction.}
	\centering
	\begin{tabular}{ l | l | l | l | l | l }
		\hline
		& $A_2^0\langle r^2 \rangle$ &$A_4^0\langle r^4 \rangle$ & $A_6^0\langle r^6 \rangle$ &  $A_6^6\langle r^6 \rangle$ & $B_{\mathrm{ex}}$ \\
		\hline
		NdCo$_5$ & -285 & -32 & 36 &  1134 & 292 \\
		TbCo$_5$ & -118 & -20 & 20 &  440 & 310 \\
		\hline
	\end{tabular}
\end{table}

\section{Crystal-field 4$f$ states and parameters in RECo$_5$} \label{app:CF_states}

In Tables~\ref{table:WFs_NdCo5} and \ref{table:WFs_TbCo5} we list the calculated 4$f$ wave functions within the GSM of Nd and Tb. The coordinate system is chosen in accordance with Ref.~\onlinecite{Alameda1982}, i.e. with the local quantization axis $z||a$ and $x||c$, where $a$ and $c$ are lattice [100] and [001] directions of the hexagonal unit cell. The states are written as the expansion $\sum a(J,M) |JM\rangle$ in pure angular momentum eigenstates $|JM\rangle$ of a given occupancy; all contributions with $a^2(J,M) > 10^{-3}$ are shown. Apart from the mixed GS in Nd and pure $|JJ\rangle$ GS state in Tb one may  also notice drastically stronger $J$-mixing effects in the case of Nd, in agreement with the significant impact of $J$-mixing on its magnetic anisotropy (Sec.~\ref{sec:T_dependence}).

For the reader's convenience we list the CFPs and $B_{\mathrm{ex}}$ in NdCo$_5$  and TbCo$_5$ calculated in the present work in Table~\ref{table:CFP_comparison}. 




\end{document}